\documentclass[a4paper,11pt,showkeys,amsmath,amssymb
]{article}
\usepackage{jheppub} 
\usepackage[T1]{fontenc} 
\usepackage{subcaption}
\usepackage[usenames,dvipsnames]{xcolor} 
\usepackage{array}
\usepackage{pifont} 
\usepackage{float} 
\usepackage{placeins} 
\definecolor{titlecolor}{rgb}{0.6,0,0}
\definecolor{mightnightblue}{RGB}{25,25,112}
\definecolor{brown}{rgb}{0.59, 0.29, 0.0}
\definecolor{darkred}{rgb}{0.6,0,0}
\definecolor{linkcolor}{rgb}{0,0,0.5}

\title{\boldmath \color{BrickRed} Democratic parameterization and analysis for 331 model as a subgroup of $SU(6)$}

\author[a]{Rena \c{C}ift\c{c}i,}
\author[b]{Abbas Kenan \c{C}ift\c{c}i,}
\author[c,d,1]{and Oleg Popov\note{Corresponding author.}}

\affiliation[a]{Department of Physics, Faculty of Science, Ege University, 35040 Bornova, Izmir, T\"{u}rkiye}
\affiliation[b]{Department of Physics, Faculty of Arts and Sciences, Izmir University of Economics, 35330 Bal\c{c}ova, Izmir, T\"{u}rkiye}
\affiliation[c]{Department of Biology, Shenzhen MSU-BIT University, 1, International University Park Road, Shenzhen 518172, China}
\affiliation[d]{Department of Physics, Korea Advanced Institute of Science and Technology, \\291 Daehak-ro, Yuseong-gu, Daejeon 34141, Republic of Korea}

\emailAdd{rena.ciftci@ege.edu.tr}
\emailAdd{kenan.ciftci@ieu.edu.tr}
\emailAdd{opopo001@ucr.edu}

\arxivnumber{2212.xxxx}
\date{\today}

\abstract{
A democratic parameterization is introduced for $SU(3)_C\otimes SU(3)_L\otimes U(1)_X$ extension of the Standard Model, which is inspired by $SU(6)$ symmetry. In the novel scenario all Cabibbo-Kobayashi-Maskawa mixing angles and quark masses, nine observable quantities in total, are predicted within 1$-$3 standard deviations of the experimental values with a minimum number of input parameters. The present work provides the thorough numerical analysis and correlations between input parameters and predicted quantities. $\chi^2\approx 0.67$ with $\forall\sigma<0.61$ corresponds to the best global fit benchmark point. Benefits of the new parameterization and future prospects are discussed as well.
}
\begin{document} 
\maketitle
\flushbottom
%
\section{Introduction}
\label{sec:intro}
Although the Standard Model (SM) is effective in accurately describing all fundamental forces excluding the gravity, it suffers from  large mass spectra and fermionic hierarchies, small quark mixing angles, and the existence of three fermion generations, violation of CP, etc. A number of  extensions to SM have been considered to address some of these issues. The so-called 331 model is one of the simplest extensions that change the electroweak gauge group of SM from $SU(3)_C \otimes SU(2)_L \otimes U(1)_Y$ to $SU(3)_C \otimes SU(3)_L\otimes U(1)_X$. Initially, these models were presented as a natural explanation for the  number of fermionic families observed in nature. 

Many research have focused on the 331 based model, which was inspired by the need to solve issues in numerous phenomenological applications. For example, papers on 331 model include, but not limited to, applications in neutrino mass generation~\cite{Boucenna:2014ela,Tully:2000kk}, flavour physics~\cite{Duy:2022qhy,Addazi:2022frt,Buras:2013dea,Buras:2012dp}, and another phenomenological challenges~\cite{Singer:1980sw,Pisano:1992bxx,Frampton:1992wt,Reig:2016tuk,Long:1995ctv,CarcamoHernandez:2005ka,Liu:1993gy,Profumo:2013sca,CarcamoHernandez:2013krw,Fonseca:2016tbn}. Beyond that, taking into account by now well recognized W boson mass anomaly, reported earlier this year by the Collider Detector at Fermilab (CDF) Collaboration obtained at Tevatron particle accelerator~\cite{CDF:2022hxs}, probable interconnection between W mass anomaly and the super-symmetric version of the 331 model has been studied~\cite{Rodriguez:2022wix}. For more up to date articles on 331 models check~\cite{CarcamoHernandez:2013zrj,CarcamoHernandez:2014wdl,CarcamoHernandez:2017cwi,Barreto:2017xix}. From another side, models based on 331 gauge group may be interpreted as a forerunner to grand unification models at high energy scales~\cite{Deppisch:2016jzl,Kownacki:2017uyq,Kownacki:2018lkj}. At last, 3311 model, an extended alternative of the 331 model, has been studied in the context with neutrino mass generation mechanism and dark matter candidates~\cite{Alves:2016fqe,Dong:2017zxo,Kang:2019sab,Leite:2019grf}.

Various variations of the 331 type have been studied in detail so far. This model can be made anomaly free in a variety of ways. Model 331  can be made anomaly free within the family like SM. Alternatively, other variants can use all three families and be anomaly free. The second approach is very attractive because it naturally explains SM's family number of three.

In a recent article submitted by authors of the present paper \cite{Ciftci:2022lrc}, Democratic Mass Matrix (DMM) approach has been applied to  $SU(3)_C\otimes SU(3)_L\otimes U(1)_X$ extension of the SM inspired by $E_6$ symmetry. The model, named as Variant-A, is anomaly free per generation of quarks and leptons~\cite{Sanchez:2001ua}. In the same work, another anomaly free model in quark/lepton generation, named as Variant-B, is given as $SU(3)_C\otimes SU(3)_L\otimes U(1)_X$ extension of the SM and is inspired by both $SU(6)$~\cite{Hartanto:2005jr} and $SU(6)\otimes U(1)_X$~\cite{Martinez:2001mu2} symmetries. While Variant-A has additional (isosinglet) quarks in down sector, Variant-B has additional (isosinglet) quarks in up sector. Details of the latter variant are given in Ref.~\cite{Ponce:2001jn}.

Introductory literature on 331 models and DMM approach  can be reached in our mentioned paper  \cite{Ciftci:2022lrc}. Yukawa coupling constants of the weak interaction Lagrangian are assumed by DMM to be about the same. Fermions acquire various masses as a result of the small deviations from the full democratic mass matrices through calculation of mass eigenvalues. The democratic parameterization of Variant-A was very successful to fit into the recent experimental data \cite{Fritzsch:2021ipb} on quark masses and CKM mixing matrix at the scale of mass of Z boson~\cite{Ciftci:2022lrc}. In this study, we want to confirm that the same parameterization works for the Variant-B of 331 model as well. Since a summary of motivation and short history of the DMM approach and 331 Model is given in our earlier paper, we will not go over it again.

The structure of the paper is as follows: The quark content and new gauge bosons, neutral and charged currents, DMM parameterization of the variation B of the 331 model are given in Section~\ref{sec:model}. Section~\ref{sec:model_param} contains details and definition of the new parameterization for the B variant. Numerical analysis and generated correlation graphs are given in section~\ref{sec:numerical_analysis}. Analysis results, more precisely the input parameters and obtained observable variables for the three most important and relevant benchmark points are presented in section~\ref{sec:results}. Future prospects and features of the collected results are discussed in Section~\ref{sec:discussion}. Conclusion is given in section~\ref{sec:conclusion}.
%
\section{331 Model}
\label{sec:model}

The model with $SU(3)_C\otimes SU(3)_L\otimes U(1)_X$ electroweak gauge group  is one of the minimal extensions of SM. It is possible to envisage various sub-models of this model with no exotic electrically charged particles~\cite{Ponce:2001jn}. Triangle anomalies can be eliminated throughout each generation. In fact, one of these models (Variant-A) was taken into consideration by the authors of this work in an earlier paper. Since this study's focus is on the prospect of democratic parameterization of another model (Variant-B) with anomalies canceling in each generation independently, a brief overview of the model's quark sector, charged currents, and neutral currents is provided in this sub-section.

\subsection{Quark content of Variant-B}

The quark structure for this model~\cite{Ponce:2001jn} is as following:

\begin{align}
\label{eq:q_rep_1}
\begin{matrix}
Q^{\alpha}_{L} = \left( \begin{array}{c}
u_{\alpha}\\
d_{\alpha}\\
U_{\alpha}
\end{array}
 \right)_{L}
& u^{c}_{\alpha L} & d^{c}_{\alpha L} & U^{c}_{\alpha L}, \\
\quad \quad \left\lbrace  3,3,\frac{1}{3}\right\rbrace  & \left\lbrace  3^{*},1,-\frac{2}{3}\right\rbrace & \left\lbrace  3^{*},1,\frac{1}{3}\right\rbrace & \left\lbrace  3^{*},1,-\frac{2}{3}\right\rbrace
\end{matrix}
\end{align}
where $ \alpha =1, 2, 3 $ correspond to the three families. Numbers in parenthesis refer to $\left( SU(3)_{C},\right.$ $SU(3)_{L}$, $\left.U(1)_{X}\right)$ quantum numbers, where $X$ arising in the electric charge generators of the gauge group is defined as
\begin{align}
    \label{eq:q_charge}
    Q = \frac{1}{2} \lambda_{3L} + \frac{1}{2\sqrt{3}} \lambda_{8L} + X I_{3},
\end{align}
where $ \lambda_{iL}$ ($i=1,\dots,8$) are Gell-Mann matrices for $ SU(3)_{L} $ and $ I_{3} $ is 3-dimensional identity matrix.
\subsection{Higgs and New Gauge Bosons}	
Model contains three Higgs fields, which are $(\phi_1^{-}, \phi_1^{0}, \phi_1^{'0})$, $(\phi_2^{-}, \phi_2^{0}, \phi_2^{'0})$ and $(\phi_3^{0}, \phi_3^{+}, \phi_3^{'+})$. Vacuum Expectation Values (VEV) of Higgs fields are the following:
\begin{center}	
\begin{equation}
	\begin{array}{c}
	 \left\langle \phi_{1} \right\rangle = (0,0,M)^{T} ,\\  
	 \left\langle \phi_{2} \right\rangle = (0,\frac{\eta}{\sqrt{2}},0)^{T},\\
	 \left\langle \phi_{3} \right\rangle = (\frac{\eta \prime}{\sqrt{2}},0,0)^{T},
	\end{array}
\end{equation}
\end{center}
where $\eta\sim 250$ GeV ($\eta \prime =\eta$ can be taken for simplicity).

Moreover, there are a total of 17 gauge bosons in this model.  One of the gauge fields is the gauge boson associated with $U(1)_X$. Eight of them are associated with $SU(3)_C$. Gauge bosons of $W^\pm$, $K^\pm$, $K^0$ and $\bar{K}^0$ are responsible from the charged current  in the electroweak sector. $Z$ and $Z^\prime$ bosons are given for neutral current, which are also massive and uncharged. The masses of the new bosons are proportional to the symmetry breaking scale of the model (order of a few TeV). The masses of the gauge bosons of the electroweak sector can be obtained with the help of  expressions below: 

\begin{subequations}
\label{eq:masses_bosons}
\begin{align}
	&m^{2}_{W^{\pm}} = \frac{g^2}{4} (\eta^{2}+\eta^{\prime 2}),\hspace{2.5cm} \\
	&m^{2}_{Z} = \frac{m^{2}_{W^{\pm}}}{C_W^2},\hspace{3.2cm}\\
	&m^{2}_{K^{\pm}} = \frac{g^2}{4} (2 M^{2}+\eta^{\prime 2}),\hspace{2.5cm} \\
	&m^{2}_{K^{0}(\bar{K}^0)} = \frac{g^2}{4} (2 M^{2}+\eta^{2}),\hspace{2.5cm} \\
	&m^{2}_{Z^\prime} = \frac{g^{2}}{4(3-4S_W^2)}\left[8C_W^2M^2+\frac{\eta^{2}}{C_W^2}+\frac{\eta^{2}(1-2S_W^2)^2}{C_W^2}\right],
\end{align}
\end{subequations}

where cosine and sine of the Weinberg angle are abbreviated by $C_W$ and $S_W$, respectively, and $S_W^2 = 0.23122$ as an experimental value. An important note is that there are five new gauge bosons beyond those of the SM. The masses of these BSM bosons can be tested in the bounds of the Large Hadron Collider (LHC) detectors. This is possible because TeV order mass values for these BSM bosons are still with an allowed parameter window. These gauge bosons' mass value constraints have been set by the non-observation of the specific kinds of the LHC events~\cite{Zyla:2020zbs}, that were expected to be detected. A more up-to-date and more strict limit of the $Z^\prime$ boson's mass is fixed at $M_{Z^\prime}~\text{\textgreater}~5.1$~TeV and $M_{Z^\prime}~\text{\textgreater}~4.6$~TeV at $95 \%$~CL. This was determined by using the most recent ATLAS~\cite{201968} and CMS data~\cite{CMS-PAS-EXO-19-019}, respectively.

In reality, the involvement of additional heavy gauge bosons~\cite{Zyla:2020zbs}, the charged ones often represented by $W^{\prime}$, is the characteristic shared by many models produced by expanding the SM.
By resonantly producing fermion or electroweak boson pairs, $W^{\prime}$ bosons would be seen in the LHC. A large amount of lost transverse energy and a high-energy electron or muon make up the most widely considered signature. At the moment, the stringent limits on the $W^{\prime}$'s mass are set at $M_{W^{\prime}}~\text{\textgreater}~6$~TeV with $95 \%$ CL~\cite{PhysRevD.100.052013}, under an assumption of the coupling between SM fermions and model's BSM gauge bosons. Despite the fact that this restriction have no direct effect on the model under consideration, nonetheless, it acts as a guide for the $K^{\pm}$ and $K^0$ bosons' mass values.

The model's Charged Currents (CC) are expressed as follows
\begin{align}
\label{eq:lag_cc}
    \mathcal{L}_{CC} &= -\frac{g}{\sqrt{2}} \left[ \bar{\nu}_{L}^\alpha \gamma^{\mu}e_{L}^{\alpha}W_{\mu}^{+}+\bar{N}_{L}^\alpha \gamma^{\mu}e_{L}^{\alpha}K_{\mu}^{+}+\bar{\nu}_{L}^\alpha \gamma^{\mu} N_{L}^{\alpha} K_{\mu}^{0} + \bar{u}_{\alpha L}\gamma^{\mu}d_{\alpha L} W_{\mu}^{+} \right. \nonumber \\
    & \left. + \Bar{U}_{\alpha L} \gamma^{\mu} d_{\alpha L}  K_{\mu}^{+} - \Bar{U}_{\alpha L} \gamma^{\mu} u_{\alpha L}  K_{\mu}^{0} + \text{h.c.} \right],
\end{align}
and neutral currents (NC) are given by
\begin{align}
\label{eq:lag_nc}
    \mathcal{L}^{NC} &= -\frac{g}{2CW} \sum_{f} \left[ \Bar{f} \gamma^{\mu} \left( g_{V}^{\prime} + g_{A}^{\prime} \gamma^5 \right) f Z_{\mu}^{\prime} \right],
\end{align}
where $f$ stands for SM quarks and leptons; $g$, $g_{V}^{\prime}$, and $g_{A}^{\prime}$ are the SM and BSM gauge coupling constants of $SU(3)_L$ symmetry's gauge bosons after its Spontaneous Symmetry Breaking(SSB).

From the above expression, we can see that $K^{\pm}$ BSM gauge bosons mediate transitions between SM down type quarks and BSM isosinglet U type quarks, where as the interactions between SM up type quarks and BSM isosinglet U type quarks are mediated by $K^0$ and $\Bar{K}^0$ BSM gauge bosons.
\subsection{Democratic Approach to the Quark Sector of 331 Model}
\label{sec:dem_quark_331}
The Democratic Mass Matrix (DMM) technique was created by H. Harari and H. Fritzsch ~\cite{HARARI1978459,FRITZSCH1979189,FRITZSCH1987391,FRITZSCH1990451,FRITZSCH1994290}to solve the issues of mass hierarchy and mixings, however it was unable to correctly predict the mass of the top quark. A number of publications were published that addressed this issue by using DMM to four families of SM~\cite{Datta,CELIKEL1995257}. ATLAS and CMS data~\cite{DJOUADI2012310,collaboration2013searches} later ruled out the SM type fourth family fermions.
As a result, if the DMM technique is right, it will invariably be applied to an extension of the SM. DMM presumes that the Yukawa coupling constants in the weak interaction Lagrangian are nearly the same. Fermions acquire distinct masses when the mass eigenstates are activated ~\cite{Atag,CIFTCI.055001,CIFTCI.053006}. 

When discussing democracy of 331 model, two different basis are defined: $ SU(3)_L\otimes U(1)_X $ symmetry basis, labeled with superscript ``$(0)$'' as in $f^{(0)}$ and the mass basis labeled without superscript as in $f$, where $f$ stands for any fermion particle. Applying the DMM technique to the Variant-B, before breaking the electroweak spontaneous symmetry (EWSS), quarks are organized as follows:

\begin{subequations}
\label{eq:eigenstates}
\begin{align}
	&\left(\begin{matrix} u^{(0)} \\ d^{(0)} \\ U^{(0)} \end{matrix}\right)_L, \quad 
	\begin{matrix}u_L^{c(0)}, & d_L^{c(0)}, & U_L^{c(0)}\end{matrix}, \\
	&\left(\begin{matrix} c^{(0)} \\ s^{(0)} \\ C^{(0)} \end{matrix}\right)_L, \quad 
	\begin{matrix}c_L^{c(0)}, & s_L^{c(0)}, & C_L^{c(0)}\end{matrix}, \\
	&\left(\begin{matrix} t^{(0)} \\ b^{(0)} \\ T^{(0)} \end{matrix}\right)_L, \quad 
	\begin{matrix}t_L^{c(0)}, & b_L^{c(0)}, & T_L^{c(0)}\end{matrix}.
\end{align}
\end{subequations}

All bases are equivalent in the case of one-family. The Lagrangian with the quark Yukawa terms for a one-family situation can be expressed as follows:

\begin{equation}
\label{eq:lag_A}
    \mathcal{L}^{Q}_{y}=Q^T_L C (a_{d}\phi_{2}d^c_L+a_u\phi_3 u^c_L+a_U\phi_1 U^c_L+a_{uU}\phi_3 U^c_L+a_{Uu} \phi_1 u^c_L)+h.c.,
\end{equation}
where $ a_{d} $, $ a_{u} $, $ a_{U} $, $ a_{uU} $ and $ a_{Uu} $ are Yukawa couplings in the $ SU(3)_{L}\otimes U(1)_{X} $ basis and $C$ is the charge conjugate operator.

In this case, we obtain a mass term for the down-quark sector:
 \begin{equation}
 m^{0}_{d}=a_{d}\frac{\eta^d}{\sqrt{2}}\quad (\eta^d= \eta^u=\eta\text{ is taken for simplicity}),
 \end{equation}
 and a mass term for the up-quark sector is given as:
  \begin{equation}
  m^{0}_{uU}= \left( \begin{array}{cc}
  a_{u} \eta^u / \sqrt{2} & \varepsilon a_{u} \eta^u / \sqrt{2} \\
 \varepsilon a_{U} \eta^U / \sqrt{2} & a_{U} \eta^U / \sqrt{2}  
  \end{array}
   \right),
\end{equation}
where $\varepsilon$ is chosen very close to one, and $ \varepsilon a_{u} $ corresponds to the $ a_{uU} $ and $ \varepsilon a_{U} $ corresponds to the $ a_{Uu}$.

To get mass eigenvalues, we need to diagonalize the mass matrix above. This is done in Ref. ~\cite{Ciftci:2016hbv} to show that this technique gives the correct $t$ and $b$ quark masses in the case of one-family.

Now, we are able to write three-family quark Yukawa Lagrangian in the $SU(3)_{L}\otimes U(1)_{X}$ basis:
\begin{align}
    \label{eq:lag_2}
    \mathcal{L}^{Q}_{y} &= \displaystyle\sum_{i=1}^{3} Q^{i T}_{L}C (a_{d}\phi_{2}d^{c}_{L}+a_{u}\phi_{3}u^{c}_{L}+a_{U}\phi_{1}U^{c}_{L}+\varepsilon a_{u} \phi_{3}U^{c}_{L}+\varepsilon a_{U}\phi_{1}u^{c}_{L}) \nonumber \\
    &+ \displaystyle\sum_{i=1}^{3} Q^{i T}_{L}C (a_{s}\phi_{2}s^{c}_{L}+a_{c}\phi_{3}c^{c}_{L}+a_{C}\phi_{1}C^{c}_{L}+\varepsilon a_{c} \phi_{3}C^{c}_{L}+\varepsilon a_{C}\phi_{1}c^{c}_{L}) \\
    &+ \displaystyle\sum_{i=1}^{3} Q^{i T}_{L}C (a_{b}\phi_{2}b^{c}_{L}+a_{t}\phi_{3}t^{c}_{L}+a_{T}\phi_{1}T^{c}_{L}+\varepsilon a_{t} \phi_{3}T^{c}_{L}+\varepsilon a_{T}\phi_{1}t^{c}_{L}) + \text{h.c.} \nonumber
\end{align}
 
%
\section{Parameterization of the Model}
\label{sec:model_param}
%
Every quark mass matrix has a little variation, symbolized by parameters labeled as $\beta$ and $\gamma$, which breaks the democratic pattern. The shape of the deviation for down, up, and heavy up BSM quarks comprises identical structure. Nevertheless, separate parameter sets are used to parameterize variances. The following are the quark mass matrices of down, up, and heavy up BSM isosinglet sectors

\begin{subequations}
\label{eq:MassM}
\begin{align}
\label{eq:MassM_u}
& \mathcal{M}^{0}_{u}=\frac{a^u \eta^u}{\sqrt{2}}\left( \begin{array}{ccc}
1+\gamma_{u} & 1  &  1-\frac{9}{2}\gamma_{u}+\beta_{u} \\
1 & 1-2\gamma_{u} & 1+3\gamma_{u}+\beta_{u} \\
 1-\frac{9}{2}\gamma_{u}+\beta_{u} & 1+3\gamma_{u}+\beta_{u} & 1+4\beta_{u} 
\end{array}
 \right), \\
\label{eq:MassM_d}
& \mathcal{M}^{0}_{d}=\frac{a^d \eta^d}{\sqrt{2}}\left( \begin{array}{ccc}
1+\gamma_{d} & 1 &  1-\frac{9}{2}\gamma_{d}+\beta_{d} \\
1 & 1-2\gamma_{d} & 1+3\gamma_{d}+\beta_{d} \\
 1-\frac{9}{2}\gamma_{d}+\beta_{d} & 1+3\gamma_{d}+\beta_{d} & 1+4\beta_{d} 
\end{array}
\right), \\
\label{eq:MassM_U}
& \mathcal{M}^{0}_{U}=\frac{a^U \eta^U}{\sqrt{2}}\left( \begin{array}{ccc}
1+\gamma_{U} & 1  &  1-\frac{9}{2}\gamma_{U}+\beta_{U} \\
1 & 1-2\gamma_{U} & 1+3\gamma_{U}+\beta_{U} \\
 1-\frac{9}{2}\gamma_{U}+\beta_{U} & 1+3\gamma_{U}+\beta_{U} & 1+4\beta_{U} 
\end{array}
\right).
\end{align}
\end{subequations}

In addition, quarks of the up sector and BSM isosinglet up quarks mix with each other, and the mixing is parameterized by $\varepsilon$ parameter according to Eq.~\eqref{eq:lag_2}:

\begin{equation}
\label{eq:MassM_uU}
    \mathcal{M}^{0}_{uU}=\left(
    \begin{array}{cc} 
        \mathcal{M}_{u}^{0} & \varepsilon_{u} \mathcal{M}_{u}^{0} \\
        \varepsilon_{u} \mathcal{M}_{U}^{0} & \mathcal{M}_{U}^{0} 
    \end{array}\right).
\end{equation}

$\mathcal{M}^{0}_{uU}$ on the $ SU(3)_{L}\otimes U(1)_{X} $ basis, $6\times 6$ mass matrix diagonalization, generates masses of up SM and Beyond Standard Model (BSM) isosinglet quarks on the mass basis.  This mass matrix can be diagonalized with the help of a $6\times 6$ unitary matrix $U_{uU}$.  While down sector quark masses are obtained by diagonalizing $\mathcal{M}^{0}_{d}$ mass matrix with a $3\times 3$ unitary matrix $U_{d}$. In a similar manner, $3\times 3$ mixing matrices, $U_{u}$ and $U_{U}$, for up type SM and heavy BSM quarks, respectively, are defined as unitary matrices that diagonalize u and U blocks of the $\mathcal{M}^{0}_{uU}$ given in Eq.~\eqref{eq:MassM_uU}. For simplicity, CP violating phases are considered to be zero from now on. As a result, diagonalizing matrices are real orthogonal matrices.

The mixing matrices $V^W_{CKM}$, $V^{K^{\pm}}$ and $V^{K^0}$ correspond to W boson of SM, whereas $K^{\pm}$ and $K^0$ are heavy BSM gauge bosons, respectively. These mixing matrices are defined through a combinations of $3\times 3$ diagonalizing matrices $U_{u}$, $U_{d}$, and $U_{U}$, mentioned above, and are given by

\begin{subequations}
\label{eq:CKM}
\begin{align}
\label{eq:CKM_SM}
& V^W_{CKM}= U_{u} U^{T}_{d} = \left( \begin{array}{ccc}
 V_{ud}  & V_{us}   & V_{ub} \\
 V_{cd}  & V_{cs}   & V_{cb} \\
 V_{td}  & V_{ts}   & V_{tb}
\end{array}
\right), \\
\label{eq:CKM_Kpm}
& V^{K^{\pm}} = U_{U} U^{T}_{d} = \left( \begin{array}{ccc}
 V_{Ud}  & V_{Us}   & V_{Ub} \\
 V_{Cd}  & V_{Cs}   & V_{Cb} \\
 V_{Td}  & V_{Ts}   & V_{Tb}
\end{array}
\right), \\
\label{eq:CKM_K0}
& V^{K^{0}} = U_{U} U^{T}_{u} = \left( \begin{array}{ccc}
 V_{Uu}  & V_{Uc}   & V_{Ut} \\
 V_{Cu}  & V_{Cc}   & V_{Ct} \\
 V_{Tu}  & V_{Tc}   & V_{Tt}
\end{array}
\right).
\end{align}
\end{subequations}

These matrices can be parameterized with three mixing angles and one phase angle:

\begin{equation}
\label{eq:CKMangles}
 V= \left( \begin{array}{ccc}
c_{12}c_{13}  & s_{12}c_{13}   & s_{13}e^{-i\delta} \\
-s_{12}c_{23}-c_{12}s_{23}s_{13}e^{i\delta}  & c_{12}c_{23}-s_{12}s_{23}s_{13}e^{i\delta}   & s_{23}c_{13} \\
s_{12}s_{23}-c_{12}c_{23}s_{13}e^{i\delta}  & -c_{12}s_{23}-s_{12}c_{23}s_{13}e^{i\delta}   & c_{23}c_{13}
\end{array}
\right),
\end{equation}
here $s_{ij}\equiv \sin\left(\theta_{ij}\right)$, $c_{ij}\equiv \cos\left(\theta_{ij}\right)$, $\theta_{ij}$ are the mixing angles, and $\delta$ is CP violating phase(not taken into account in the present work).
%
\section{Numerical analysis}
\label{sec:numerical_analysis}

The analysis performed over the model parameterization can be divided into three stages: a systematic scan over all, seven in total (for details see Tab.~\ref{tab:benchmark_param}), input parameters of the model, next a more fine grained scan near the points with minimal deviation from the experimental data is performed, then the obtained results were used as a input data for the neural network (NN) training, and further, for obtaining a complete scan over input parameter range. Following the numerical scans, the correlation analyses between different input parameters, distinctive input parameters and predicted observable variables, as well as between various output observable variables was performed. The purpose of studying these correlations is to increase the predictive power of the model and assist in probing the model in the current and future phenomenological experiments. Included below are some of the most important and relevant correlations between input parameters and/or observable variables. The task of the present section is to study the origin behind these correlations.

Results shown below were obtained with the following values for $a$ and $\eta$ (defined in Sec.~\ref{sec:dem_quark_331}) parameters
\begin{subequations}
\label{eq:a_eta_values}
\begin{align}
\label{eq:a_eta_values_u}
    \frac{a^{u} \eta^{u}}{\sqrt{2}} &= 2400~\text{GeV}, \\
\label{eq:a_eta_values_ud}
    \frac{a^{d} \eta^{d}}{\sqrt{2}} &= 0.91~\text{GeV}, \\
\label{eq:a_eta_values_D}
    \frac{a^U \eta^U}{\sqrt{2}} &= 2.4\times 10^4~\text{GeV}.
\end{align}
\end{subequations}
\begin{figure}[H]
    \centering
    \begin{subfigure}[t]{0.45\textwidth}
        \includegraphics[width=\textwidth]{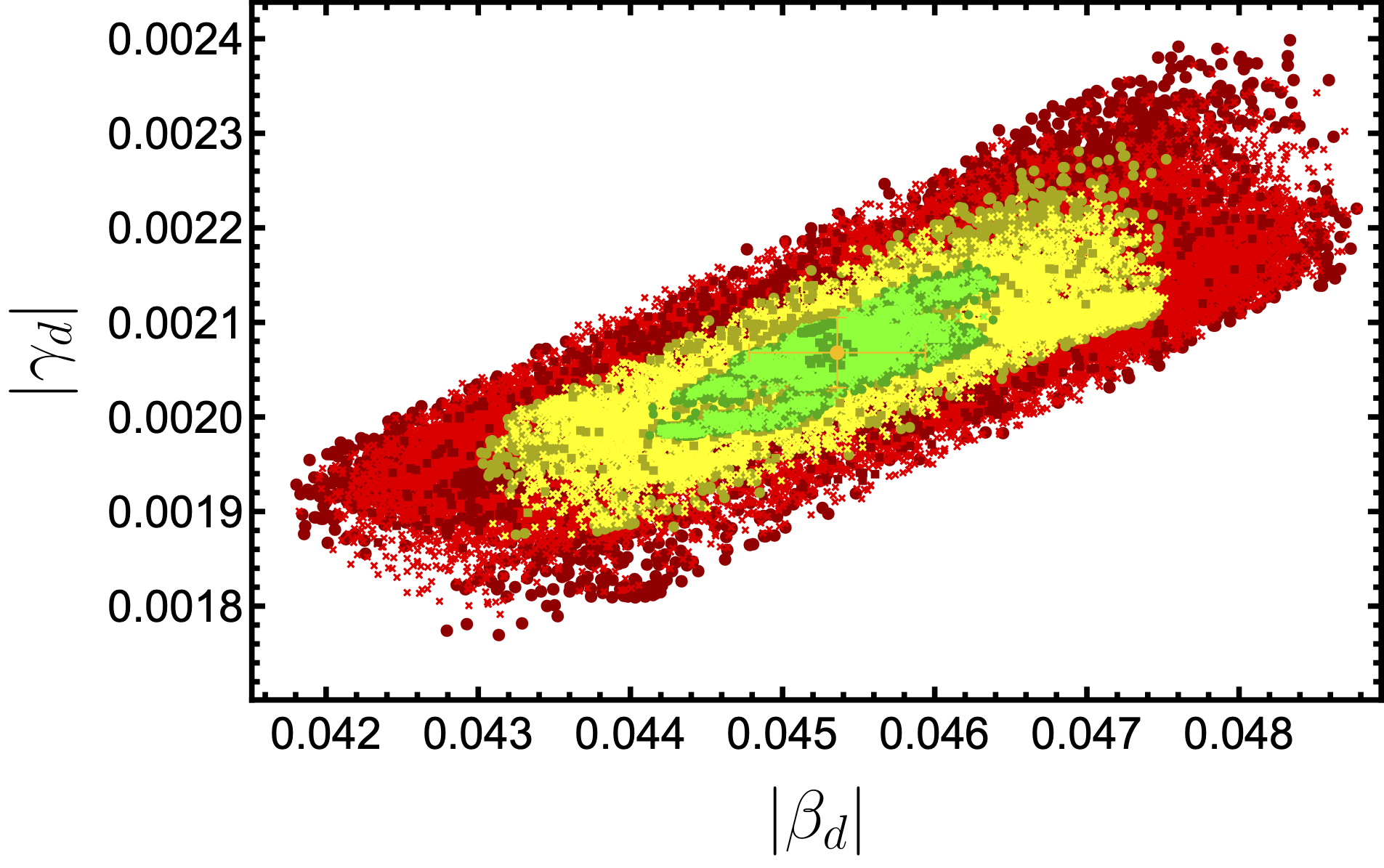}
        \caption{$\beta_d$ vs $\gamma_d$ correlation plot.}
        \label{fig:bd_gd_B}
    \end{subfigure}
    \begin{subfigure}[t]{0.47\textwidth}
        \includegraphics[width=\textwidth]{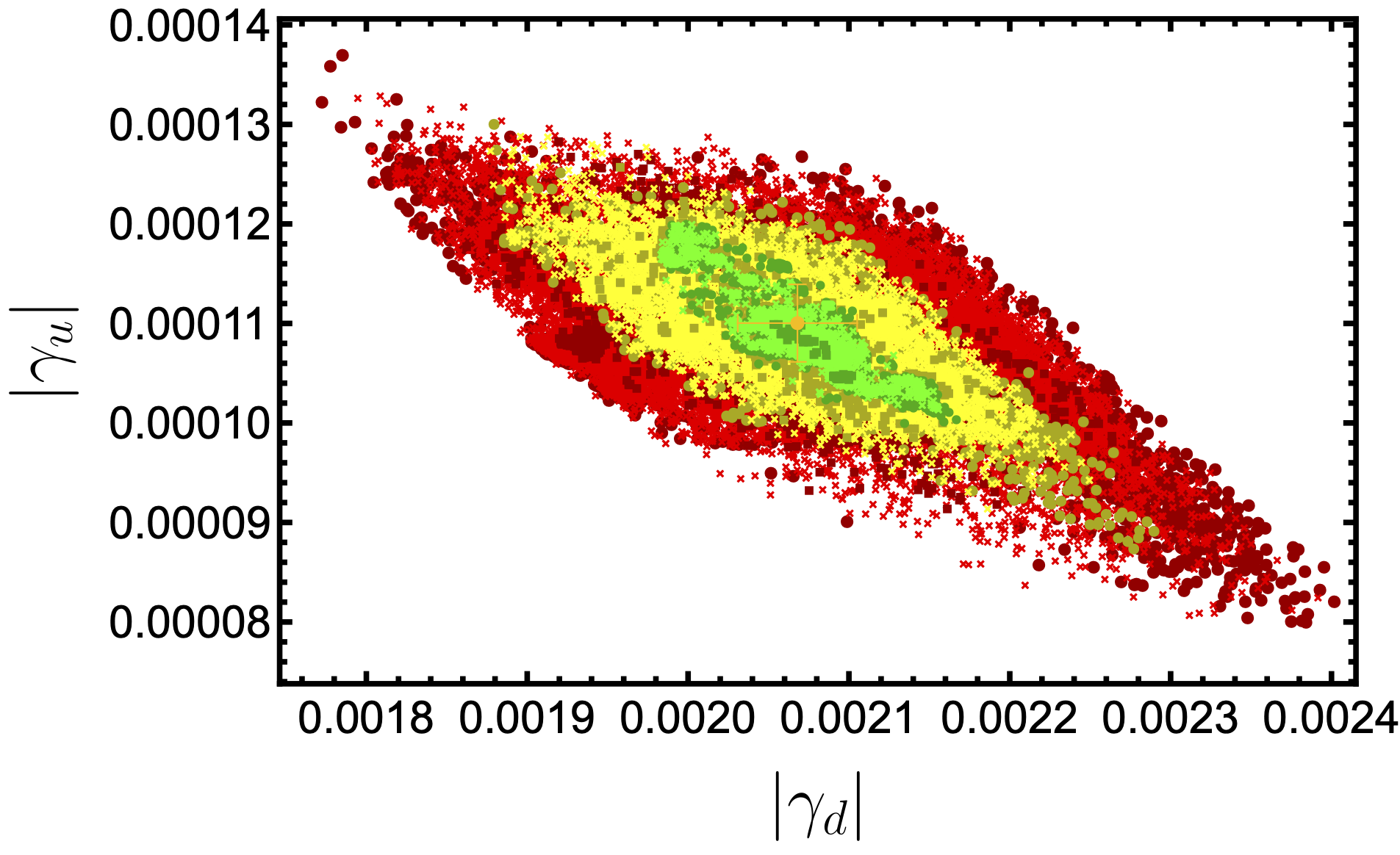}
        \caption{$\gamma_d$ vs $\gamma_u$ correlation plot.}
        \label{fig:gd_gu_B}
    \end{subfigure}
    \caption{Selected input correlation plots. Maximum standard deviation from experimental values is represented by colors. Red, yellow, and green colors are used for the values $\sigma_{\text{max}}<3$, $2$, and $1$, respectively. Whereas, discs, crosses, and squares correspond to $\left\langle\sigma\right\rangle / \sigma_{\text{max}}$: $0.5-1.0$, $0.33-0.5$, $0-0.33$, respectively.}
    \label{fig:bg_corr_B}
\end{figure}

The strongest correlation patterns between distinct input parameters are shown in Fig.~\ref{fig:bg_corr_B}. As can be seen from Fig.~\ref{fig:bd_gd_B}, there is direct correlation between $\beta_d$ and $\gamma_d$ input parameters. Fig.~\ref{fig:gd_gu_B} demonstrates the correlation between $\gamma_u$ and $\gamma_d$, which exhibits an inverse correlation contrary to the $\beta_d$ vs $\gamma_d$ case. Both of these behaviours a drastically different from analogous correlation for the Variant A of the 331 model~\cite{Ciftci:2022lrc}. Other combinations of input parameters exhibit no apparent correlation.

\begin{figure}[H]
    \centering
    \begin{subfigure}[t]{0.46\textwidth}
        \includegraphics[width=\textwidth]{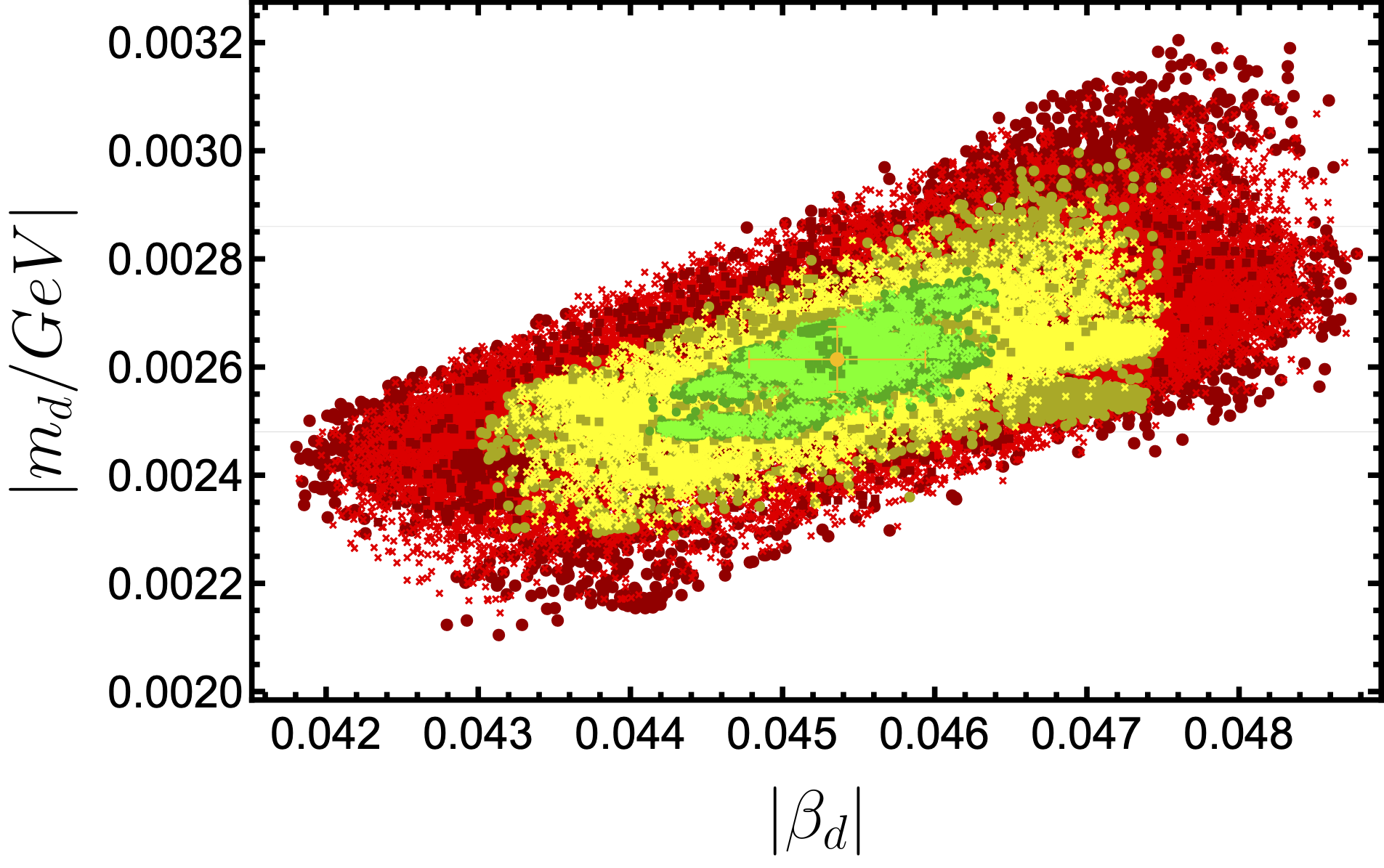}
        \caption{$\beta_d$ vs $m_d$ correlation plot.}
        \label{fig:bd_md_B}
    \end{subfigure}
    \begin{subfigure}[t]{0.45\textwidth}
        \includegraphics[width=\textwidth]{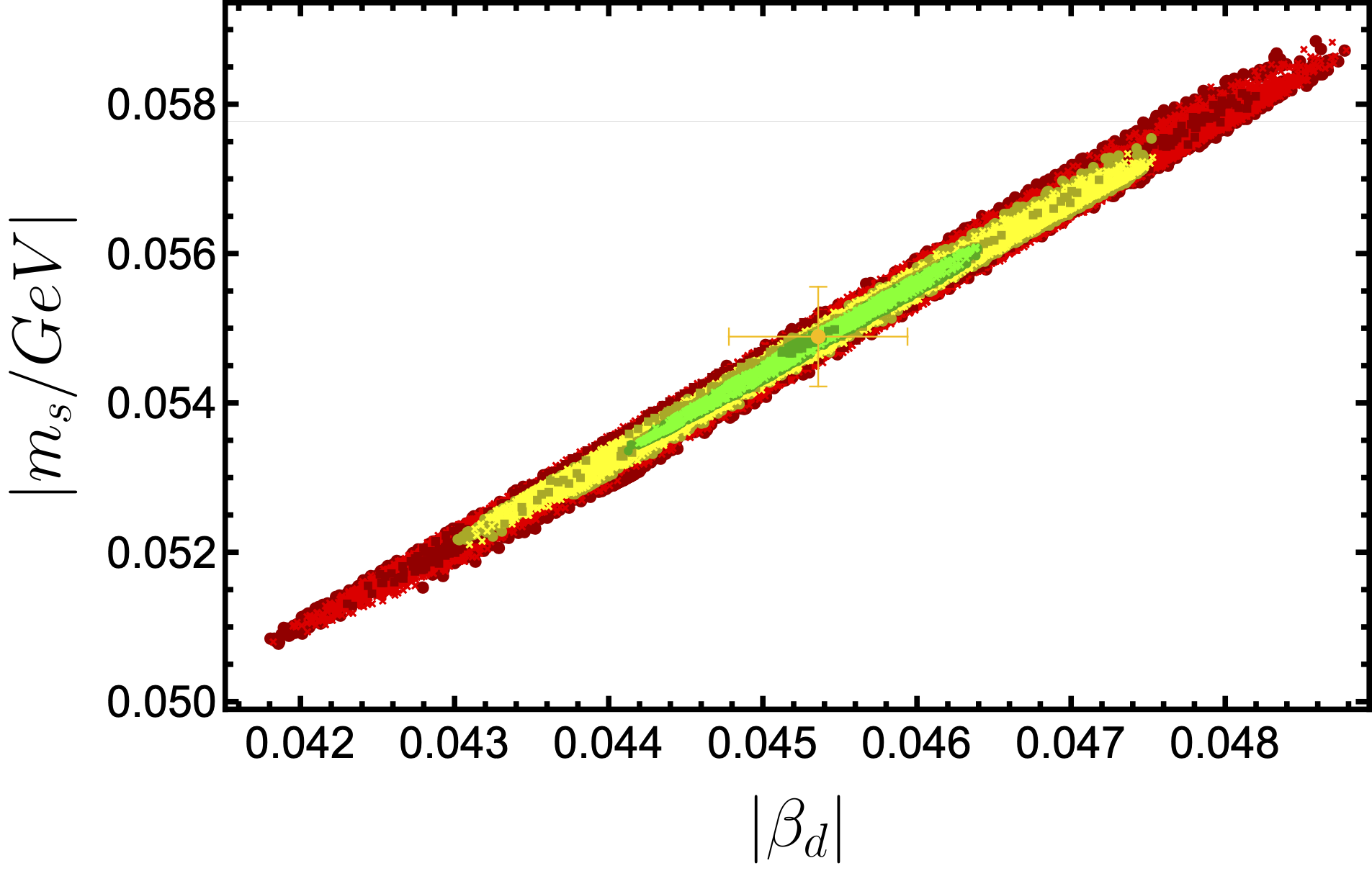}
        \caption{$\beta_d$ vs $m_s$ correlation plot.}
        \label{fig:bd_ms_B}
    \end{subfigure}
    \\
    \begin{subfigure}[t]{0.45\textwidth}
        \includegraphics[width=\textwidth]{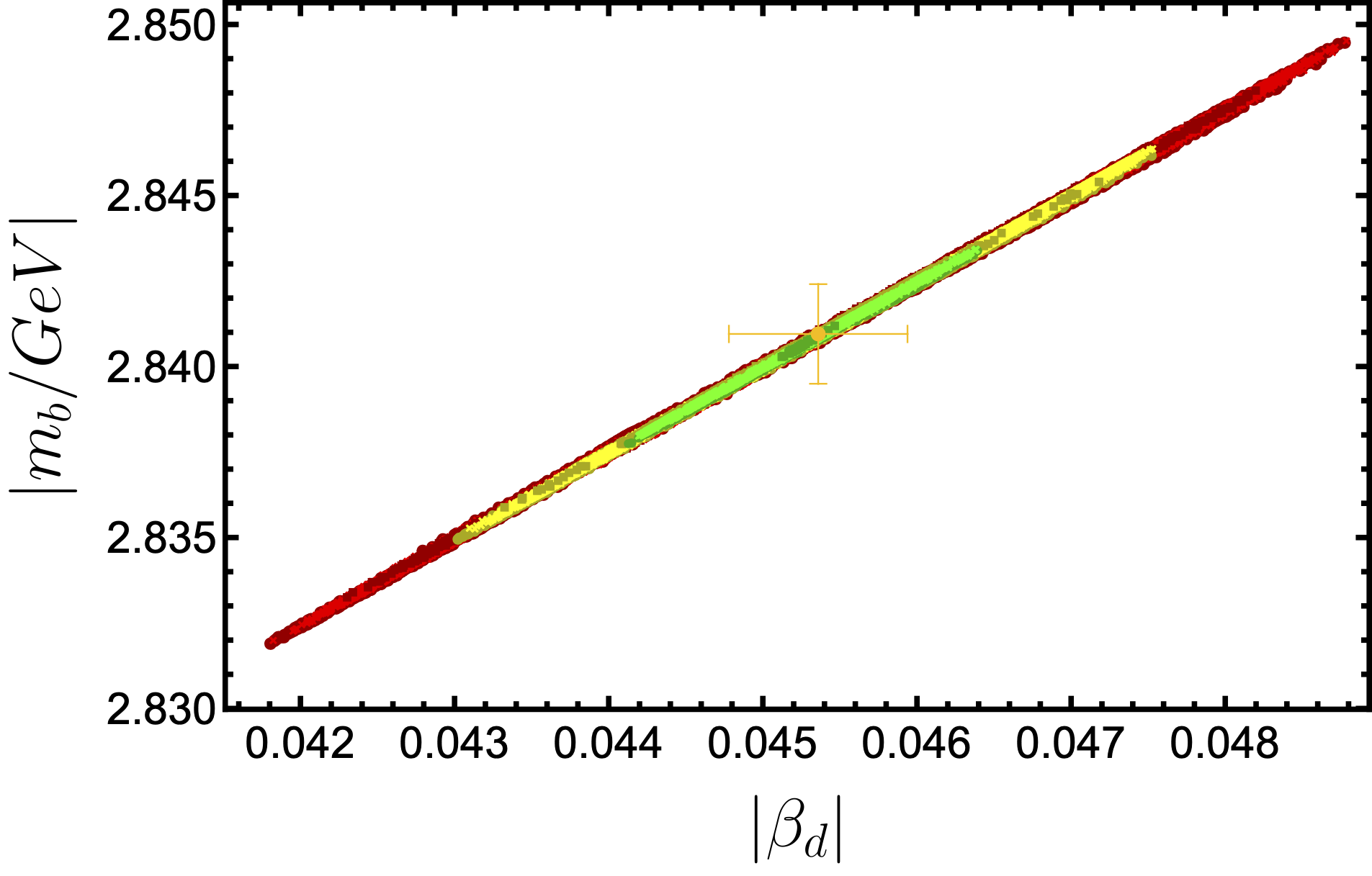}
        \caption{$\beta_d$ vs $m_b$ correlation plot.}
        \label{fig:bd_mb_B}
    \end{subfigure}
    \begin{subfigure}[t]{0.45\textwidth}
        \includegraphics[width=\textwidth]{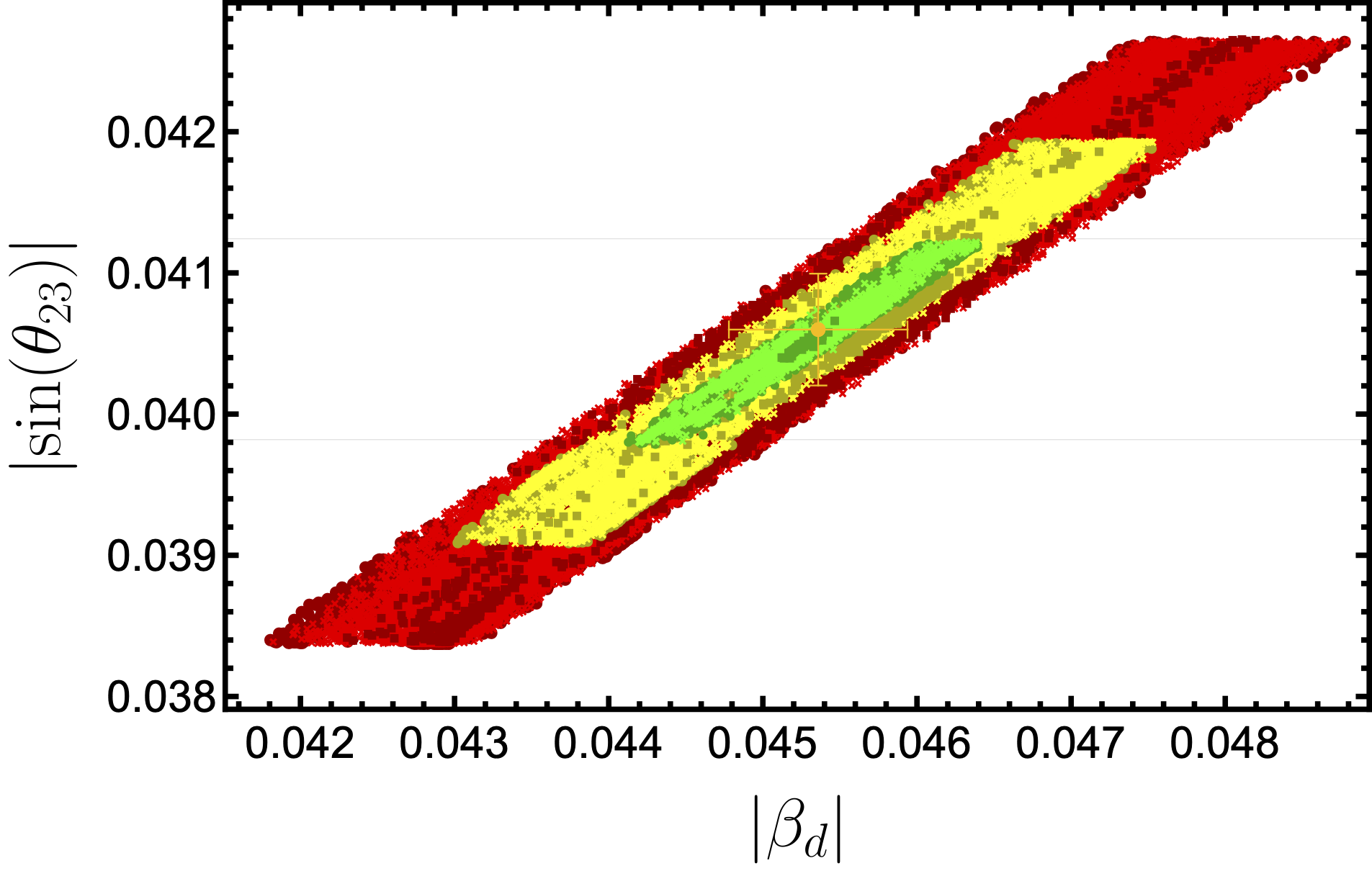}
        \caption{$\beta_d$ vs $\sin\left(\theta_{23}^{\text{\tiny{CKM}}}\right)$ correlation plot.}
        \label{fig:bd_s23_B}
    \end{subfigure}
    \caption{continued.}
\end{figure}
\begin{figure}[H]\ContinuedFloat
\centering
    \begin{subfigure}[t]{0.46\textwidth}
        \includegraphics[width=\textwidth]{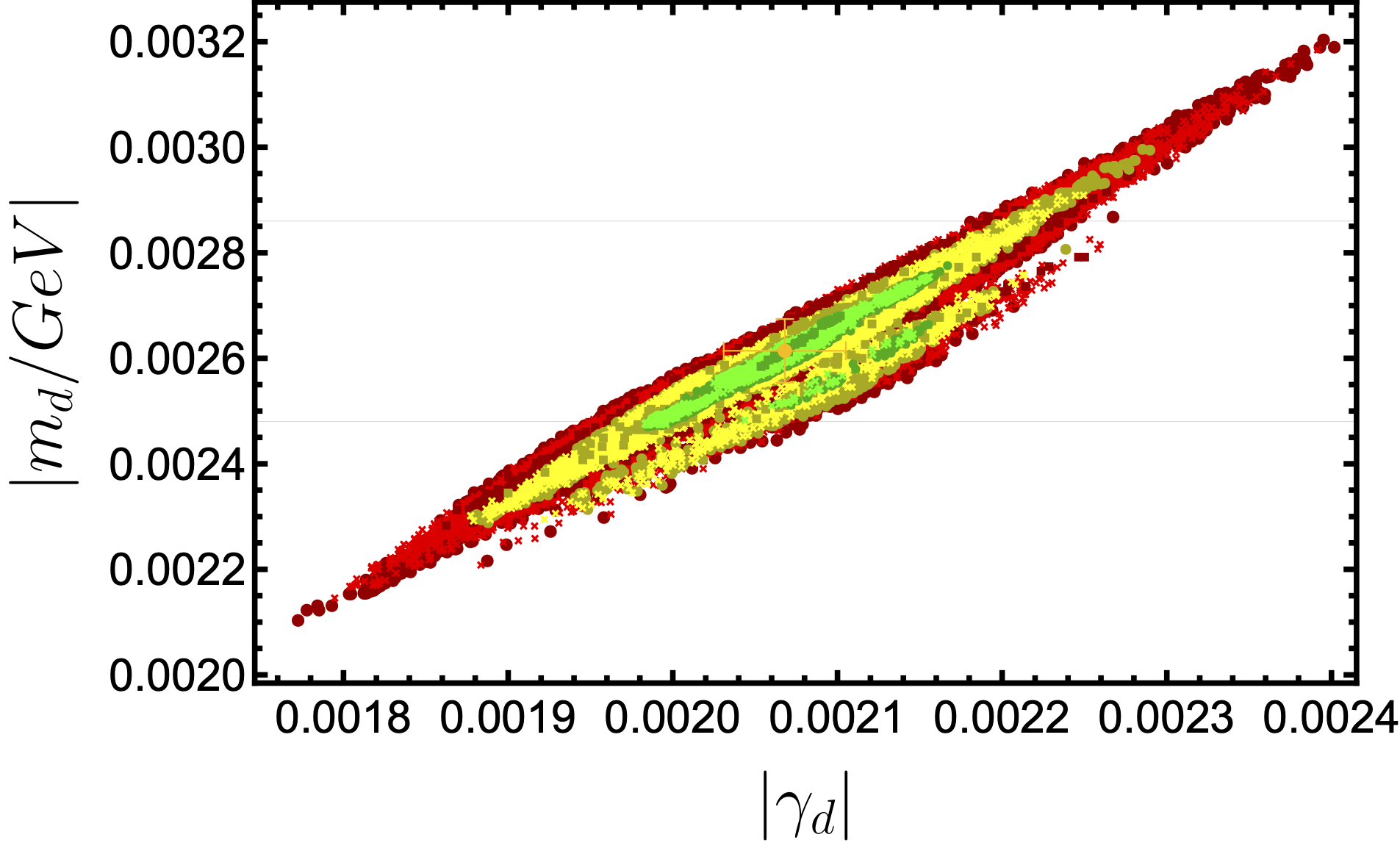}
        \caption{$\gamma_d$ vs $m_d$ correlation plot.}
        \label{fig:gd_md_B}
    \end{subfigure}
    \begin{subfigure}[t]{0.45\textwidth}
        \includegraphics[width=\textwidth]{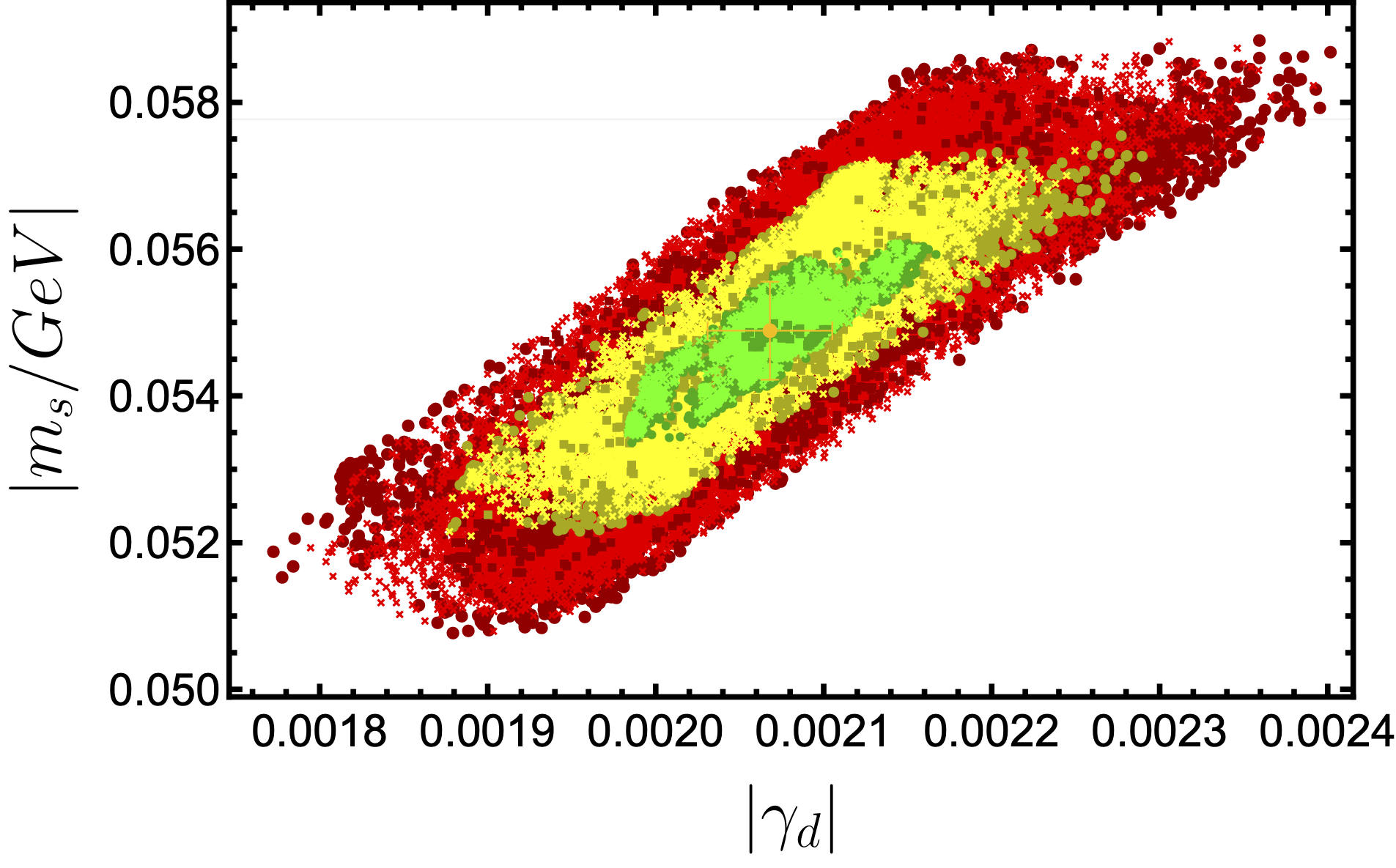}
        \caption{$\gamma_d$ vs $m_s$ correlation plot.}
        \label{fig:gd_ms_B}
    \end{subfigure}
    \\
    \begin{subfigure}[t]{0.45\textwidth}
        \includegraphics[width=\textwidth]{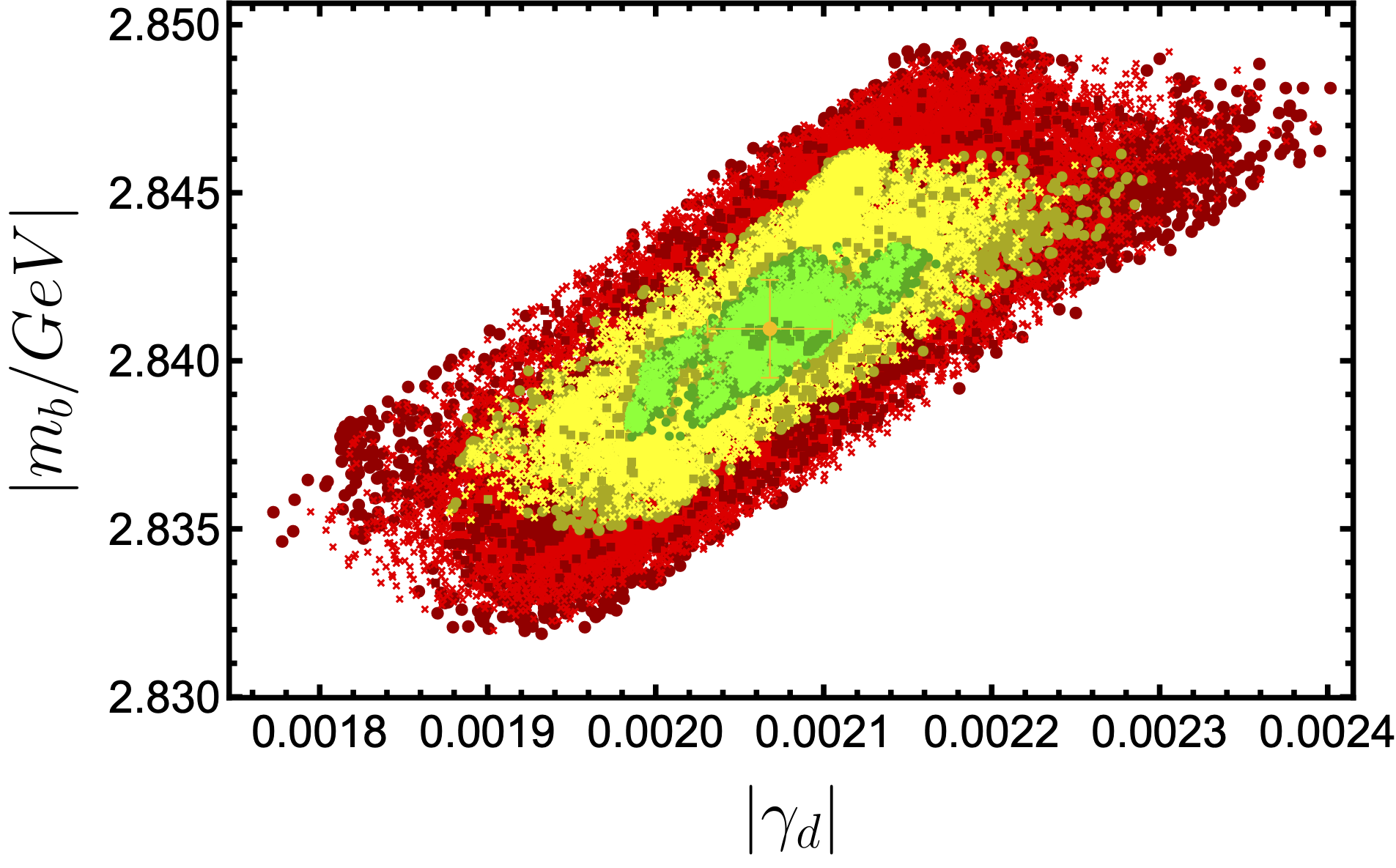}
        \caption{$\gamma_d$ vs $m_b$ correlation plot.}
        \label{fig:gd_mb_B}
    \end{subfigure}
    \begin{subfigure}[t]{0.45\textwidth}
        \includegraphics[width=\textwidth]{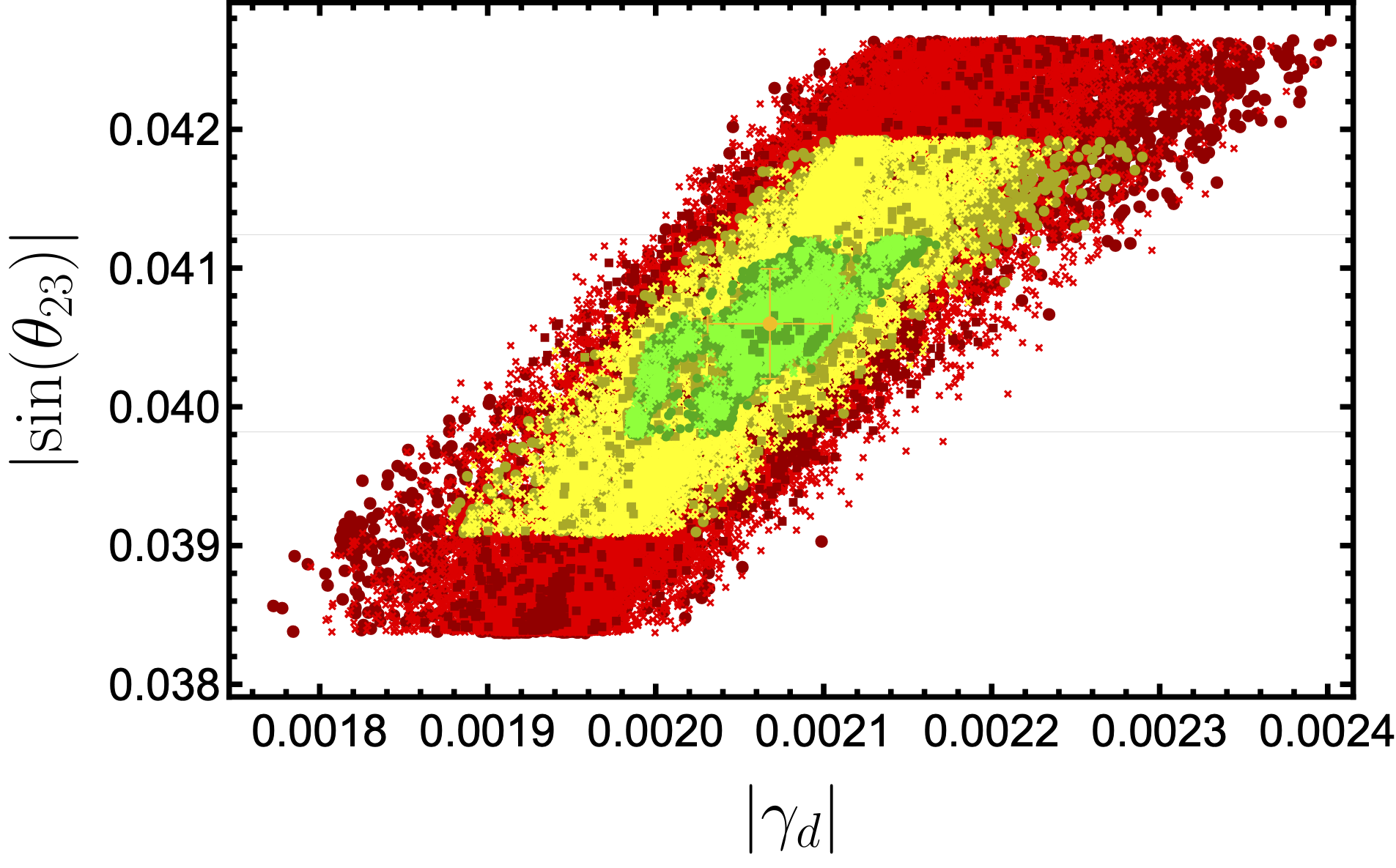}
        \caption{$\gamma_d$ vs $\sin\left(\theta_{23}^{\text{\tiny{CKM}}}\right)$ correlation plot.}
        \label{fig:gd_s23_B}
    \end{subfigure}
\end{figure}
\begin{figure}[H]\ContinuedFloat
\centering
    \begin{subfigure}[t]{0.45\textwidth}
        \includegraphics[width=\textwidth]{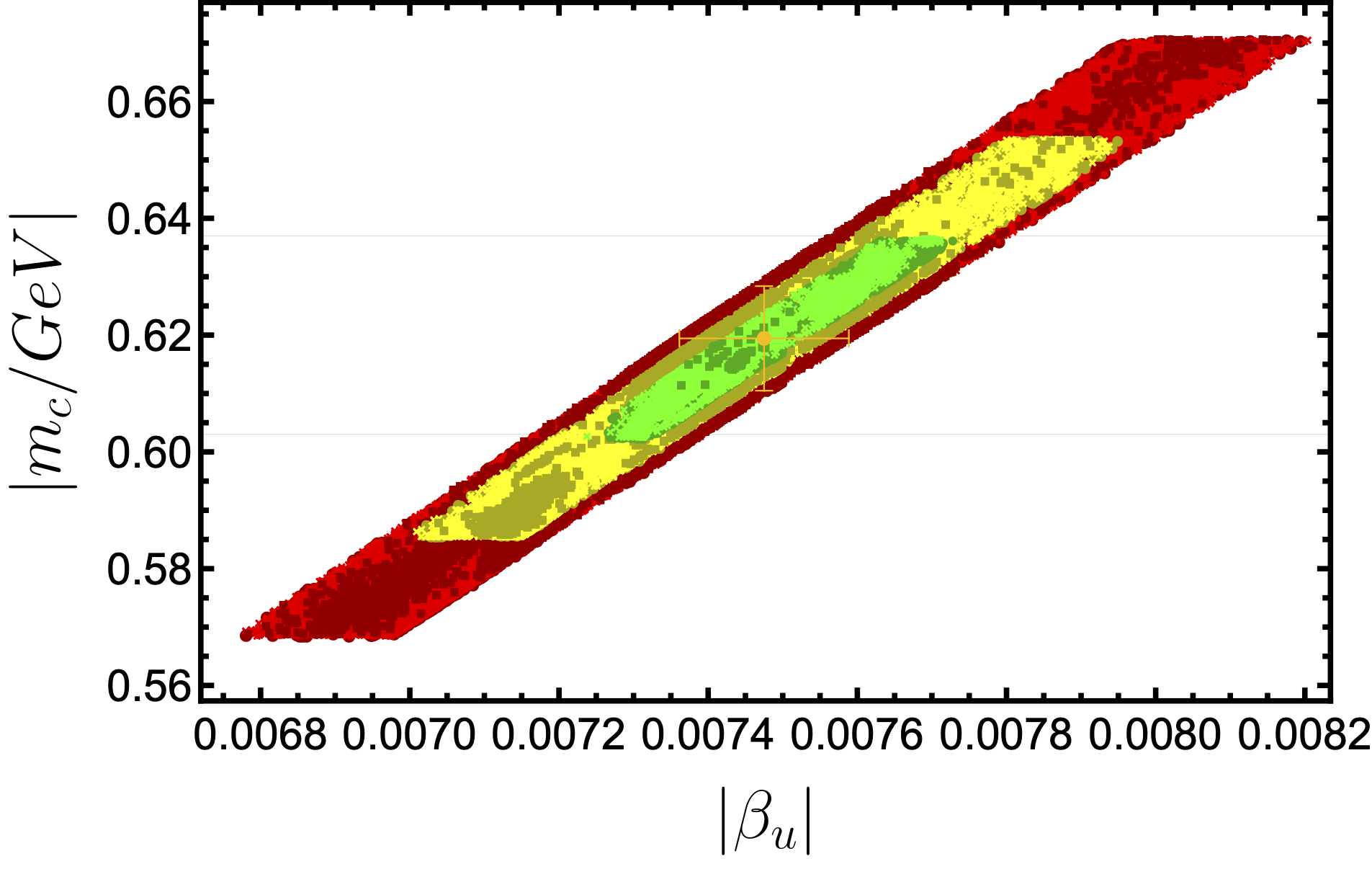}
        \caption{$\beta_u$ vs $m_c$ correlation plot.}
        \label{fig:bu_mc_B}
    \end{subfigure}
    \begin{subfigure}[t]{0.45\textwidth}
        \includegraphics[width=\textwidth]{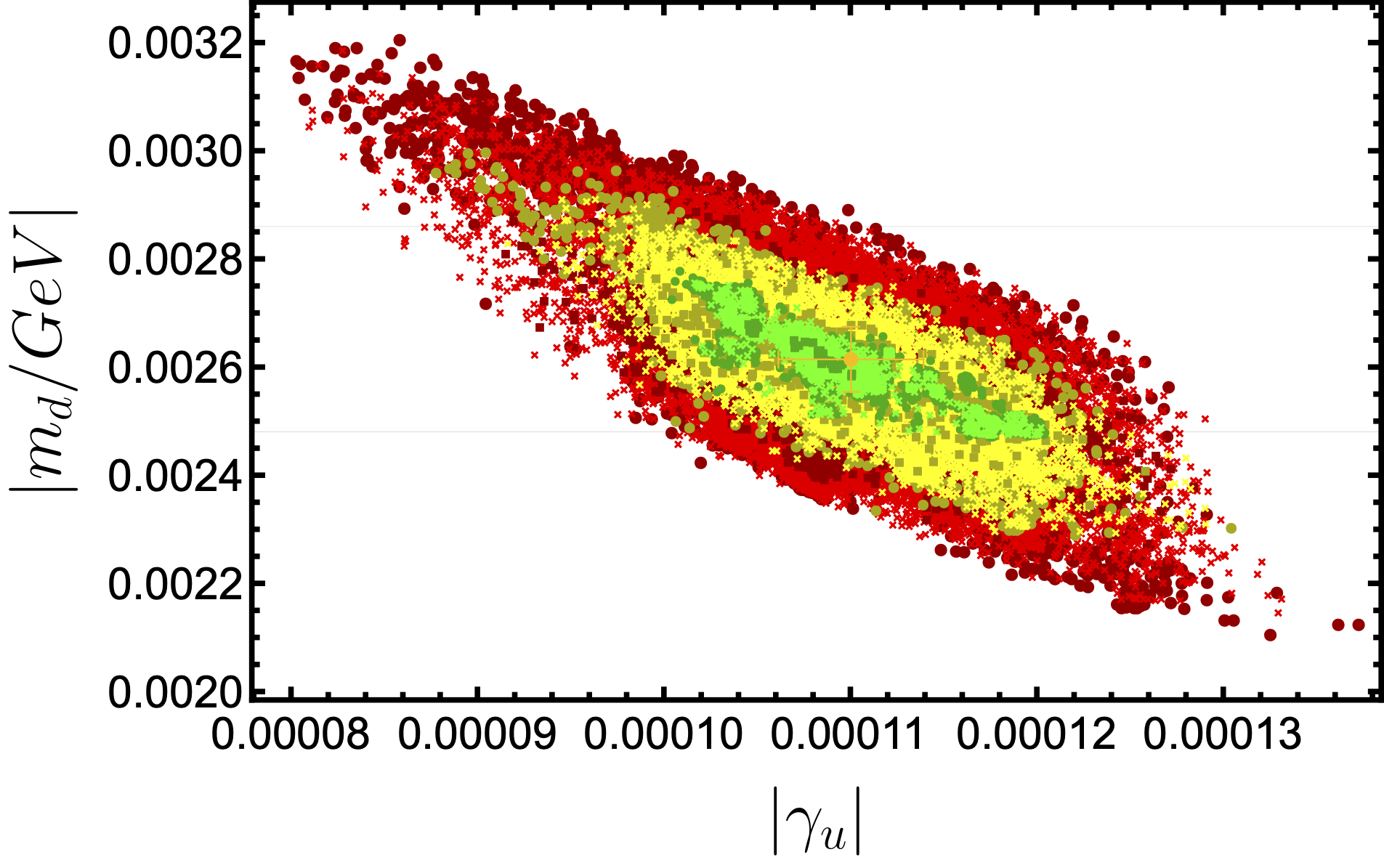}
        \caption{$\gamma_u$ vs $m_d$ correlation plot.}
        \label{fig:gu_md_B}
    \end{subfigure}
    \\
    \begin{subfigure}[t]{0.45\textwidth}
        \includegraphics[width=\textwidth]{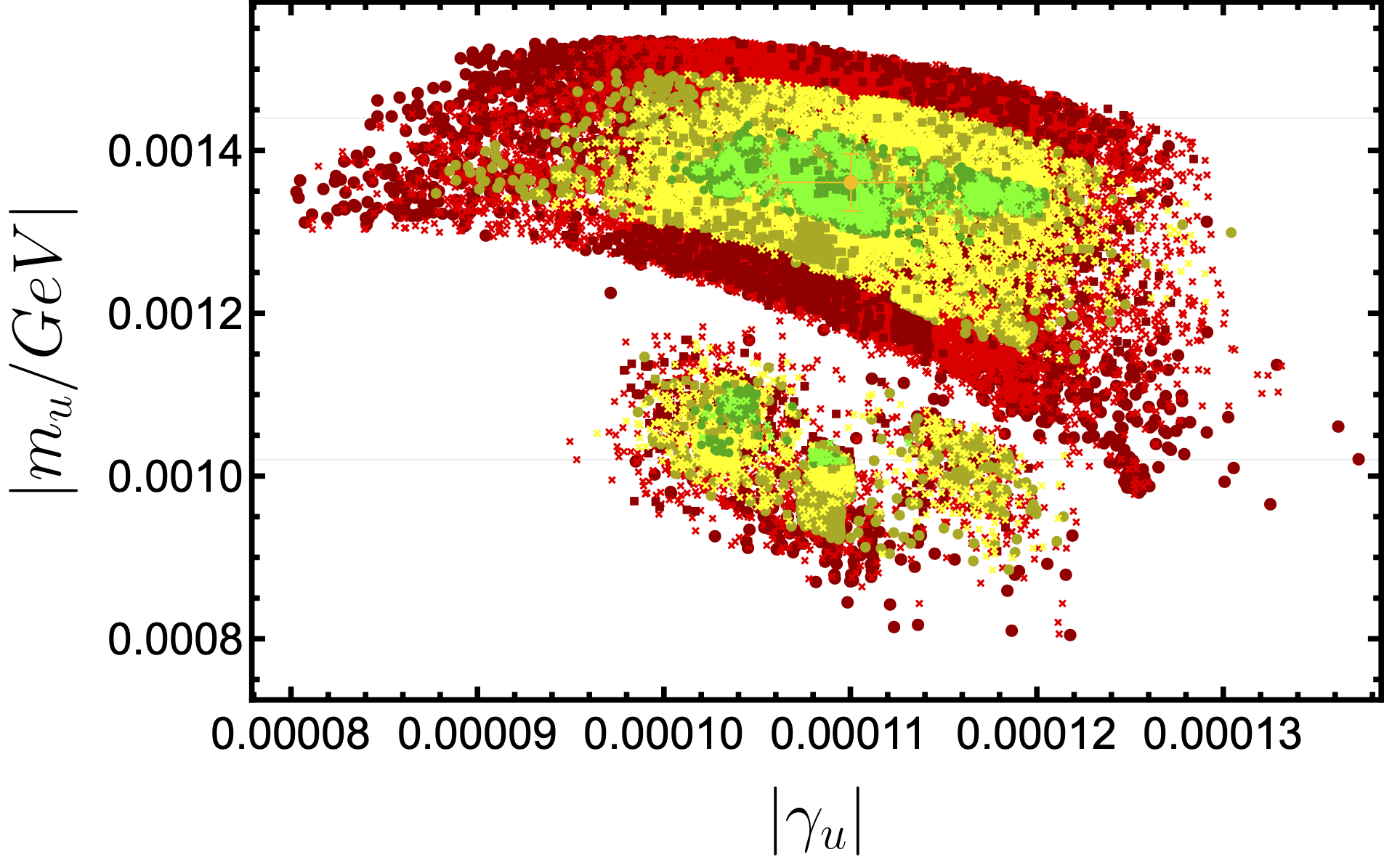}
        \caption{$\gamma_u$ vs $m_u$ correlation plot.}
        \label{fig:gu_mu_B}
    \end{subfigure}
    \caption{Selected correlation plots between input parameters and observable variables. Grid lines represent areas of the experimental data with one standard deviation. Red, yellow, and green colors are used for the values $\sigma_{\text{max}}<3$, $2$, and $1$, respectively. Whereas, discs, crosses, and squares correspond to $\left\langle\sigma\right\rangle / \sigma_{\text{max}}$: $0.5-1.0$, $0.33-0.5$, $0-0.33$, respectively.}
    \label{fig:bg_observ_corr_B}
\end{figure}

The plots in the Fig.~\ref{fig:bg_observ_corr_B} demonstrate important dependence of some observable variables on model input parameters. For instance, plot in Fig.~\ref{fig:bd_md_B} shows the direct but weak dependence of $m_d$, lightest eigenvalue of the down quark sector, on the input parameter $\beta_d$, compared to the cases with its heavier counterparts of the sector. Strong direct correlations are observed between $m_s$ and $\beta_d$, as well as between $m_b$ and $\beta_d$ (Fig.~\ref{fig:bd_ms_B},~\ref{fig:bd_mb_B}). Furthermore, from Fig.~\ref{fig:bd_s23_B} one can see that there is a similar linear dependence of $\sin\left(\theta_{23}^{\text{\tiny{CKM}}}\right)$ on $\beta_d$. An analogous correlation can be seen between $\gamma_d-m_d$, $\gamma_d-m_s$, $\gamma_d-m_b$ and $\gamma_d-\sin\left(\theta_{23}^{\text{\tiny{CKM}}}\right)$ (Fig.~\ref{fig:gd_md_B}, ~\ref{fig:gd_ms_B}, ~\ref{fig:gd_mb_B} and Fig.~\ref{fig:gd_s23_B}), which exhibit proportional direct-linear behaviour. The direct proportionality between $m_d$ and $\gamma_d$ can be seen from Eq.~\eqref{eq:MassM_d}, for which the lightest eigenvalue ($m_d$) approaches to zero as $\gamma_d$ goes to zero. Since $m_s$ and $m_b$ are most sensitive to the $\beta_d$ their $\beta_d$ plots are much thinner compared to their plots vs $\gamma_d$. Situation with $m_d$ is reversed because $\gamma_d$ has a leading effect on it. Similar to the situation in the down sector Fig.~\ref{fig:bd_ms_B}, $\beta_u$ has a strongest influence on the $m_c$, Fig.~\ref{fig:bu_mc_B}, with a direct-linear behaviour. Furthermore, an inverse proportionality between $m_d$ and $\gamma_u$, which is depicted in Fig.~\ref{fig:gu_md_B}, is originated from indirect relation between up and down sectors through the CKM mixing angles. 

From the above analysis it can be concluded that $\beta_d$ and $\gamma_d$ have noticeable influence on all the down sector mass eigenvalues and $\sin\left(\theta_{23}^{\text{\tiny{CKM}}}\right)$, whereas $\beta_u$ affects up sector quark masses. The correlation between $m_t$ and $\beta_u$ is absent due to the mixing of SM up quark sector with BSM heavy quarks. $\gamma_u$ has the strongest effect on $m_u$. Fig.~\ref{fig:gu_mu_B} demonstrates two minima of $m_u$ with respect to $\gamma_u$. Similarly, two minima are observed in other input parameters vs $m_u$ plots. $m_u$ dependence on $\gamma_u$ is not direct nor linear, unlike the situation with $\gamma_d$ and $m_d$ (Fig.~\ref{fig:gd_md_B}), this is caused by the affect of $\varepsilon$ and mixing of $m_u$ with heavy $m_U$ state. Figs.~\ref{fig:gd_md_B} and ~\ref{fig:gu_mu_B} contain similar patterns in a sense that patterns consist of two disconnected minimal regions.

Among remarkable correlation patterns of the mixing angles of the CKM matrix are expressed in the $\sin\left(\theta_{23}^{\text{\tiny{CKM}}}\right)$ mixing angle, proportional linearly and totally constrained by input parameters $\gamma_d$ and $\beta_d$ (Fig.~\ref{fig:gd_s23_B} and Fig.~\ref{fig:bd_s23_B}). Input variables demonstrate more complicated effect on the other CKM mixing angles and therefore will be skipped in the further discussion.

\begin{figure}[H]
    \centering
    \begin{subfigure}[t]{0.45\textwidth}
        \includegraphics[width=\textwidth]{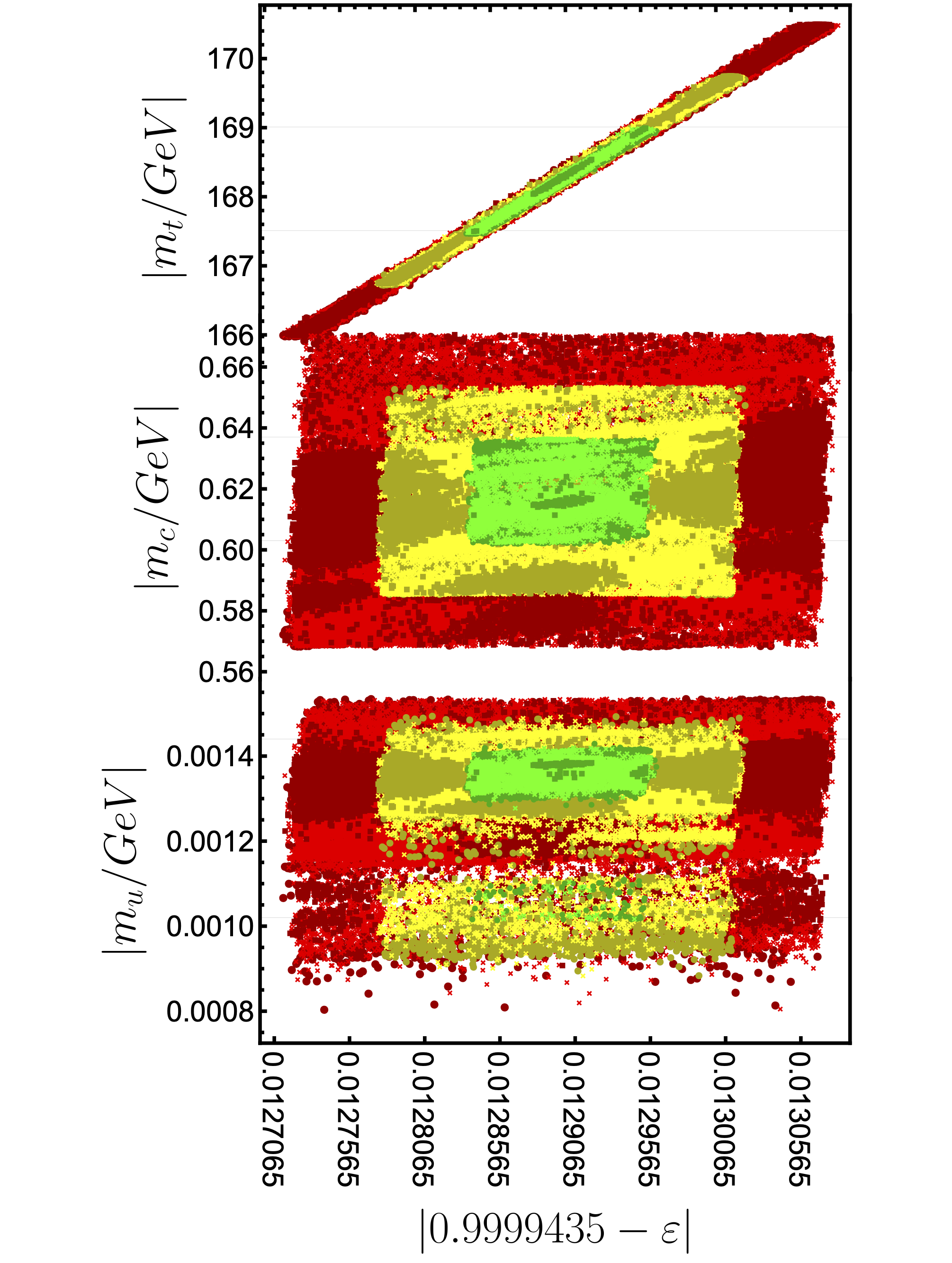}
        \caption{$\varepsilon$ vs $m_{u,c,t}$ graphs.}
        \label{fig:eps_mu_B}
    \end{subfigure}
    \begin{subfigure}[t]{0.435\textwidth}
        \includegraphics[width=\textwidth]{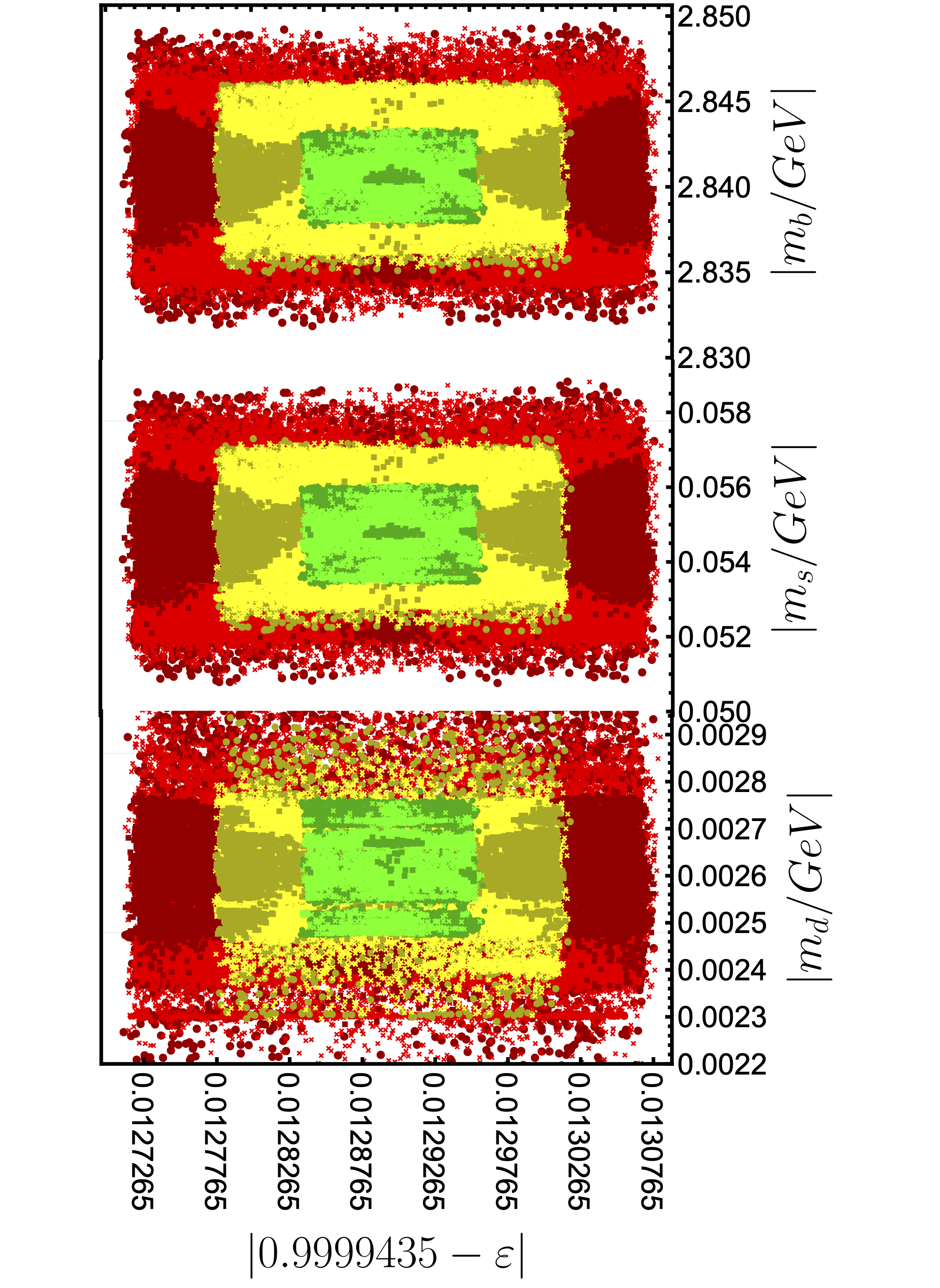}
        \caption{$\varepsilon$ vs $m_{d,s,b}$ graphs.}
        \label{fig:eps_md_B}
    \end{subfigure}
    \\
    \begin{subfigure}[t]{0.99\textwidth}
        \includegraphics[width=\textwidth]{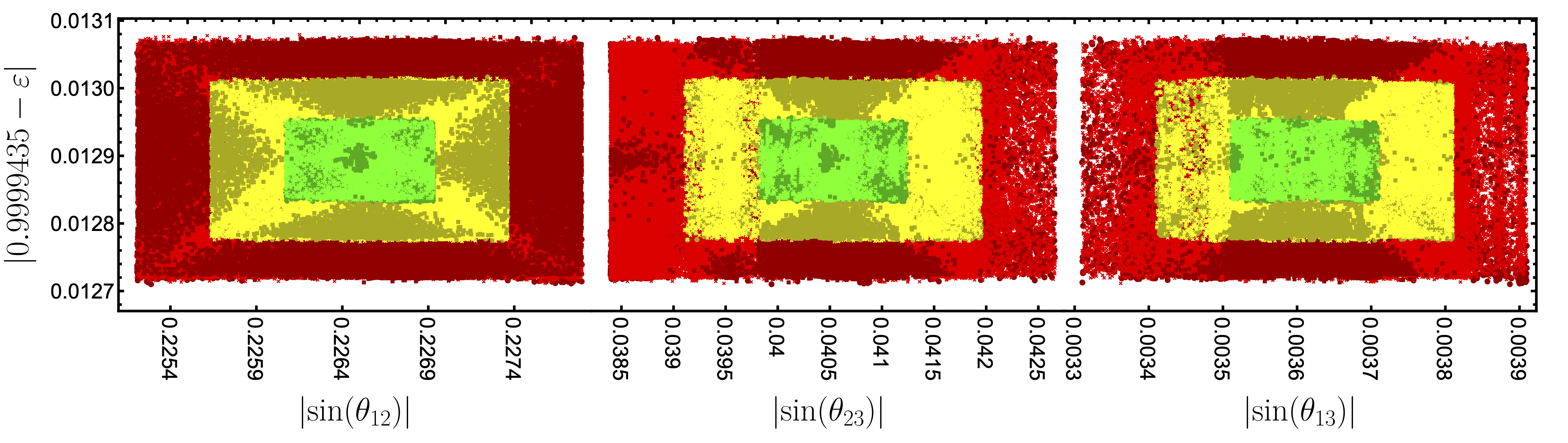}
        \caption{$\sin\left(\theta_{12}^{\text{\tiny{CKM}}}\right)$, $\sin\left(\theta_{23}^{\text{\tiny{CKM}}}\right)$, and $\sin\left(\theta_{13}^{\text{\tiny{CKM}}}\right)$ vs $\varepsilon$ graphs.}
        \label{fig:eps_s123_B}
    \end{subfigure}
    \caption{Correlation plots for $\varepsilon$ input parameter vs SM mass eigenvalues. Grid lines represent areas of the experimental data with one standard deviation. Red, yellow, and green colors are used for the values $\sigma_{\text{max}}<3$, $2$, and $1$, respectively. Whereas, discs, crosses, and squares correspond to $\left\langle\sigma\right\rangle / \sigma_{\text{max}}$: $0.5-1.0$, $0.33-0.5$, $0-0.33$, respectively.}
    \label{fig:eps_observ_corr_B}
\end{figure}
SM fermion masses and CKM mixing angles dependence on the mixing parameter($\varepsilon$) between SM up quarks and heavy BSM counterparts is depicted in Fig.~\ref{fig:eps_observ_corr_B}. $\varepsilon$ has a unique effect on the mass eigenvalues. For the up sector, the lightest mass eigenvalue, $m_u$, dominantly depends on the $\gamma_u$, whereas dependence of $m_c$ is lead by $\beta_u$, and top quark mass, $m_t$, is linearly dependent on $\varepsilon$, Figs.~\ref{fig:eps_mu_B}, in the neighborhood of $\varepsilon\rightarrow 1$ and approaches zero in its limit. This can be easily seen when $\beta_u, \gamma_u\rightarrow0$ and taking limit $\varepsilon\rightarrow 1$

\begin{subequations}
\label{eq:mt_mT_masses}
\label{eq:eps_limit}
\begin{align}
    m_t &= \left(3.96 - \sqrt{10.5+5.2\varepsilon^2}\right)\times10^4 \text{GeV}, \\
    m_T &= \left(3.96 + \sqrt{10.5+5.2\varepsilon^2}\right)\times10^4 \text{GeV},
\end{align}
\label{eq:eps_limit_2}
\begin{align}
    m_t &= -1.31\times 10^4 (\varepsilon-1) \text{GeV},\\
    m_T &= 7.92 \times 10^4 + 1.31\times 10^4 (\varepsilon-1) \text{GeV}.
\end{align}
\end{subequations}

Since the $\varepsilon$ parameter only controls the mixing in the up sector, the down sector practically is independent of it, Figs.~\ref{fig:eps_md_B}. The only indirect effect can be observed via the CKM mixing. Finally, for the CKM mixing angle dependence on $\varepsilon$ one can observe indirect relation via the mixing matrix in the up sector. This can be explained in the following way, since top quark mass, $m_t$, strongly depends on $\varepsilon$, changing which will alter the top quark mass, it will then also shift the values of mixing angles between first two and third family in the up sector, which in return will reveal itself in the CKM mixing matrix, Fig.~\ref{fig:eps_s123_B}.
\begin{figure}[H]
    \centering
    \begin{subfigure}[t]{0.45\textwidth}
        \includegraphics[width=\textwidth]{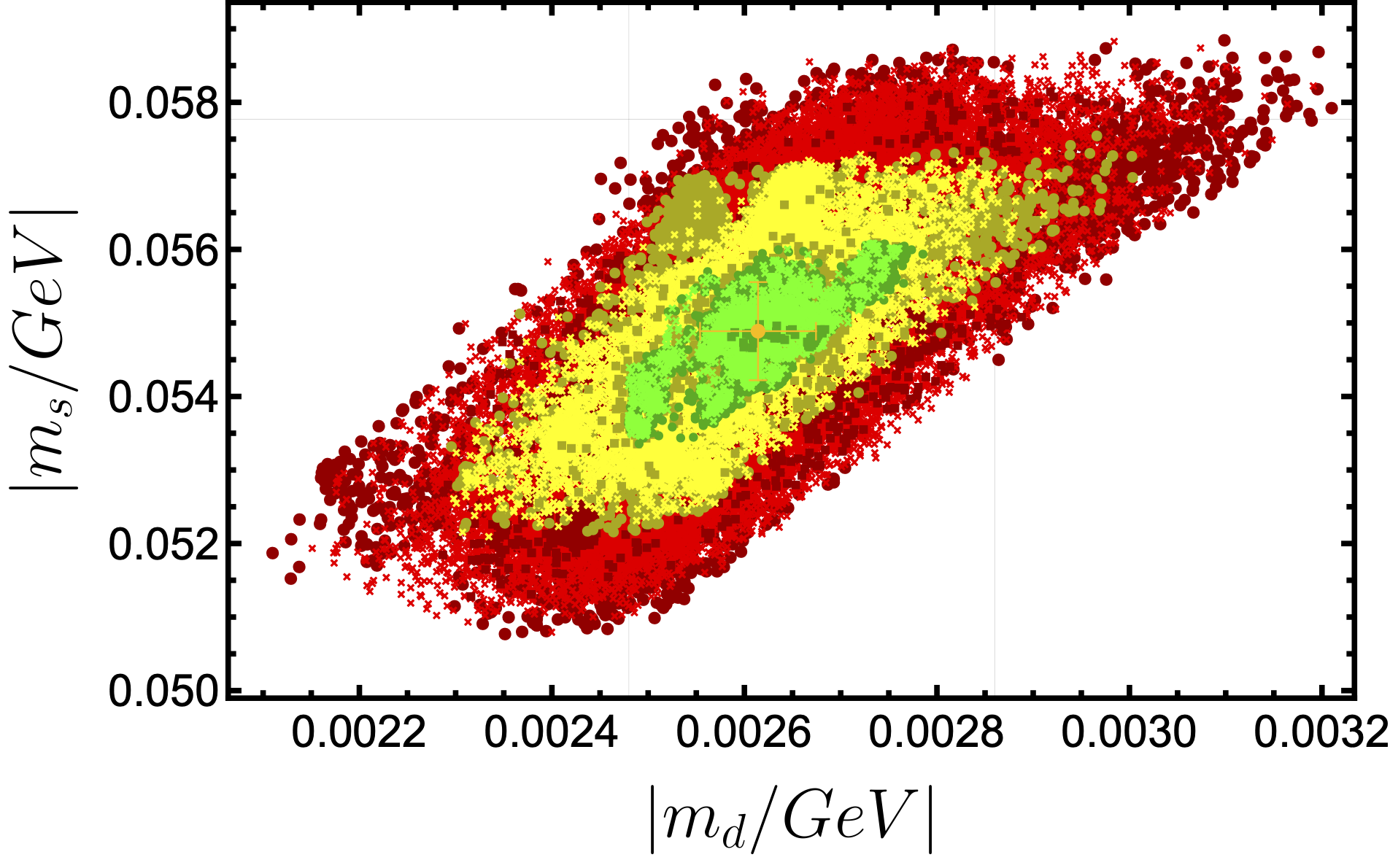}
        \caption{$m_d$ vs $m_s$ global fit distribution graph.}
        \label{fig:md_ms_B}
    \end{subfigure}
    \begin{subfigure}[t]{0.45\textwidth}
        \includegraphics[width=\textwidth]{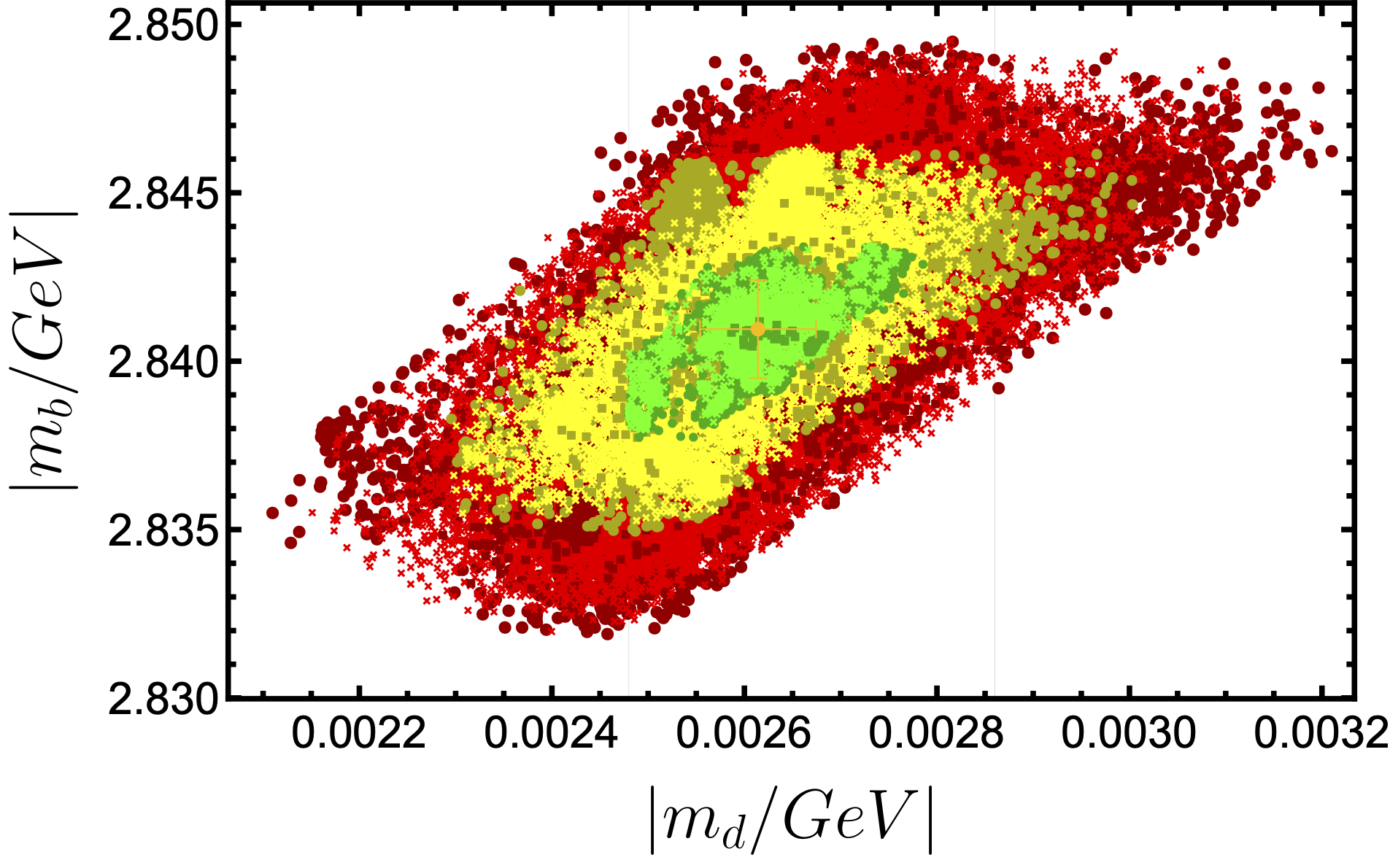}
        \caption{$m_d$ vs $m_b$ global fit distribution graph.}
        \label{fig:md_mb_B}
    \end{subfigure}
    \\
     \begin{subfigure}[t]{0.45\textwidth}
        \includegraphics[width=\textwidth]{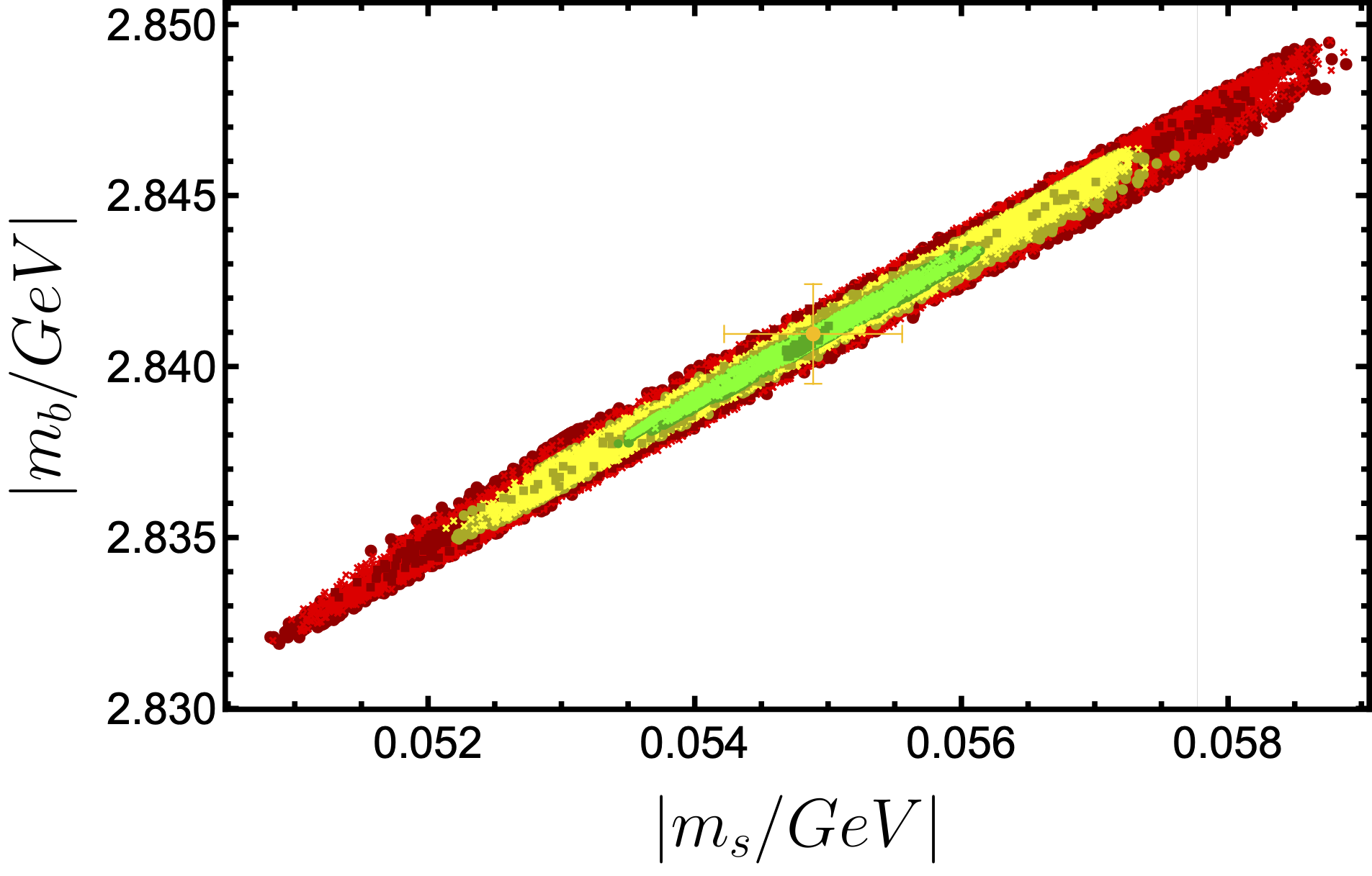}
        \caption{$m_s$ vs $m_b$ global fit distribution graph.}
        \label{fig:ms_mb_B}
    \end{subfigure}
    \begin{subfigure}[t]{0.47\textwidth}
        \includegraphics[width=\textwidth]{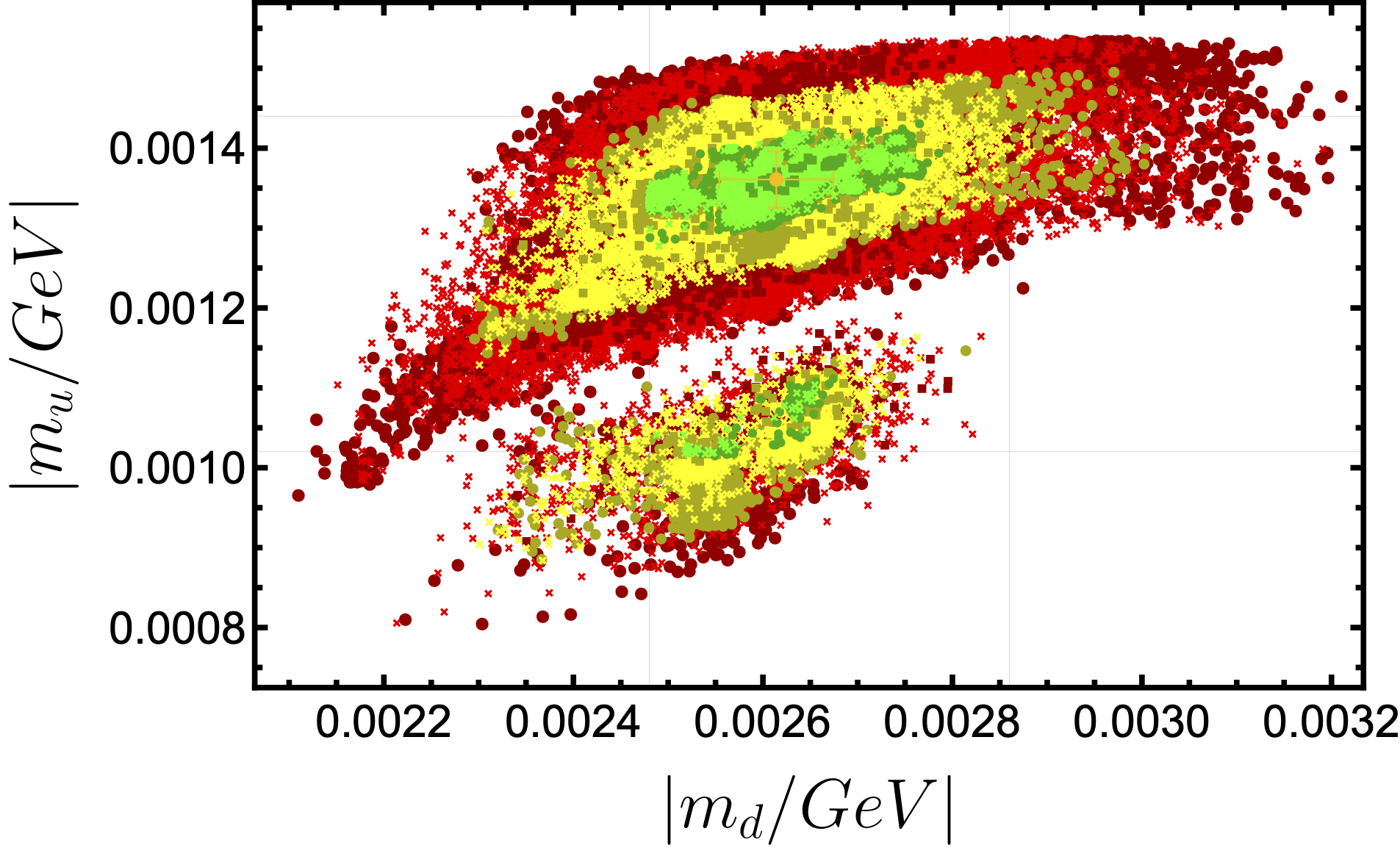}
        \caption{$m_d$ vs $m_u$ global fit distribution graph.}
        \label{fig:md_mu_B}
    \end{subfigure}
\end{figure}
\begin{figure}[H]\ContinuedFloat
    \centering
    \begin{subfigure}[t]{0.46\textwidth}
        \includegraphics[width=\textwidth]{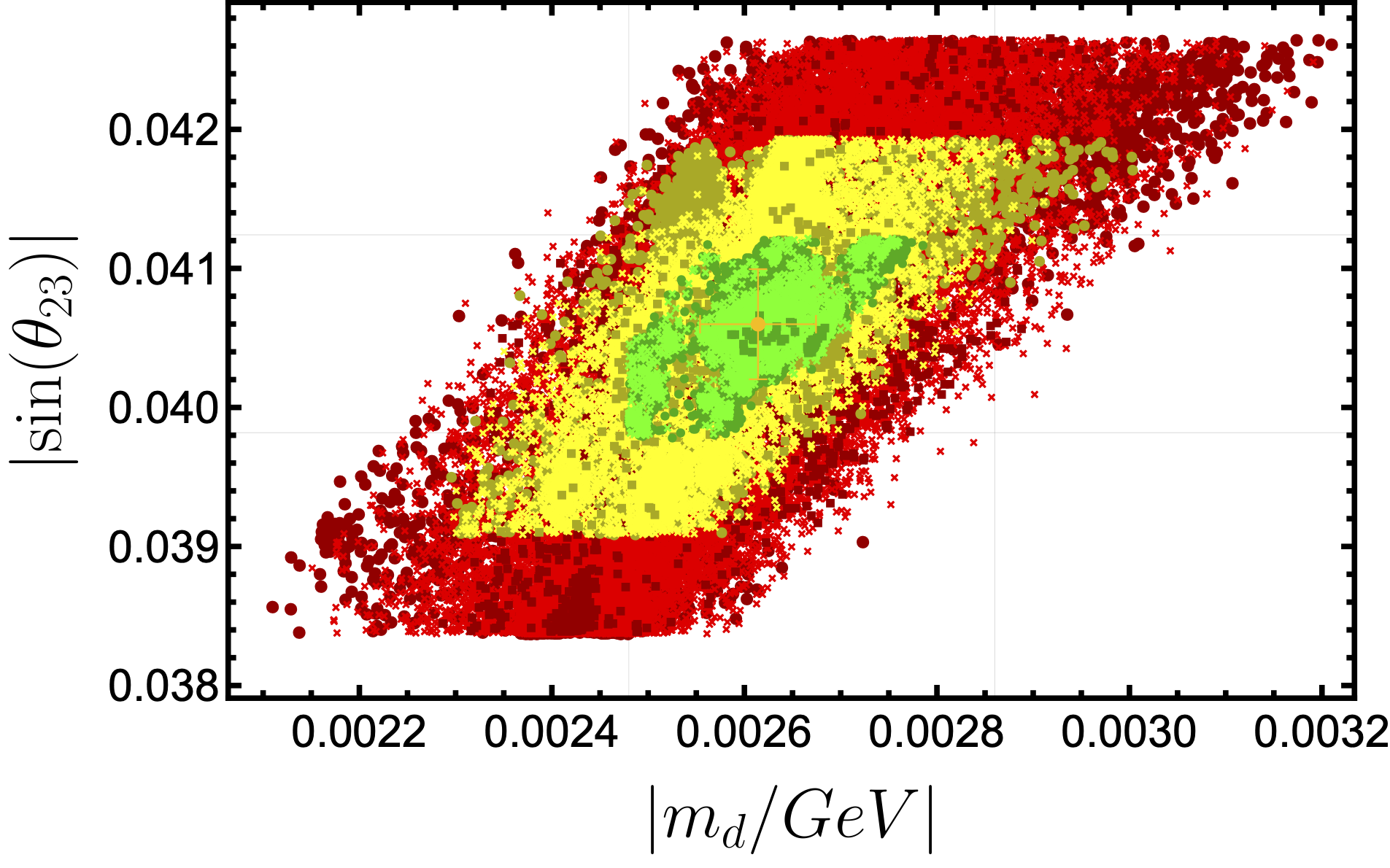}
        \caption{$m_d$ vs $\sin \left(\theta_{23}^{\text{\tiny{CKM}}}\right)$ global fit distribution graph.}
        \label{fig:md_s23_B}
    \end{subfigure}
    \begin{subfigure}[t]{0.45\textwidth}
        \includegraphics[width=\textwidth]{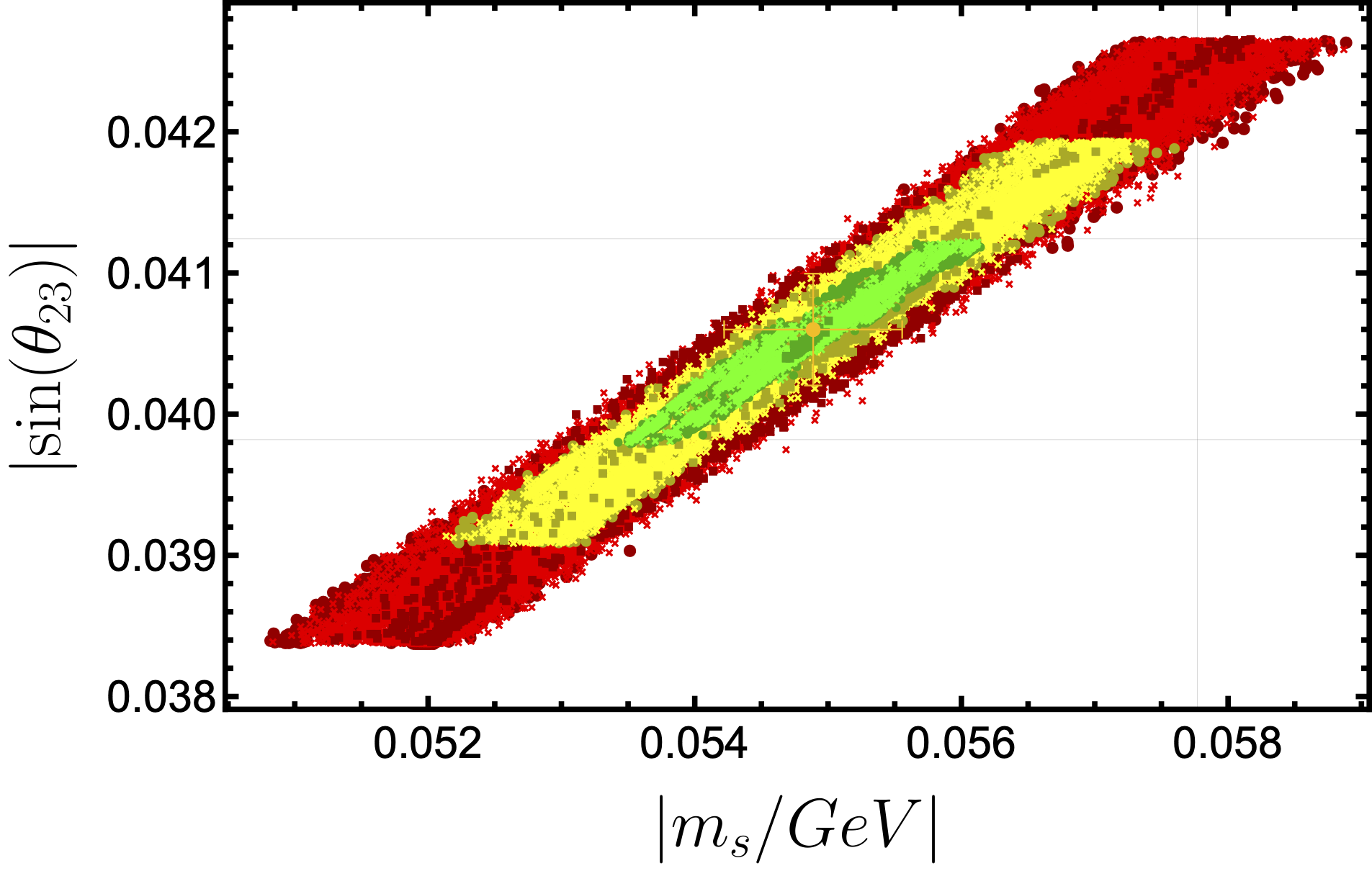}
        \caption{$m_s$ vs $\sin \left(\theta_{23}^{\text{\tiny{CKM}}}\right)$ global fit distribution graph.}
        \label{fig:ms_s23_B}
    \end{subfigure}
    \\
    \begin{subfigure}[t]{0.45\textwidth}
        \includegraphics[width=\textwidth]{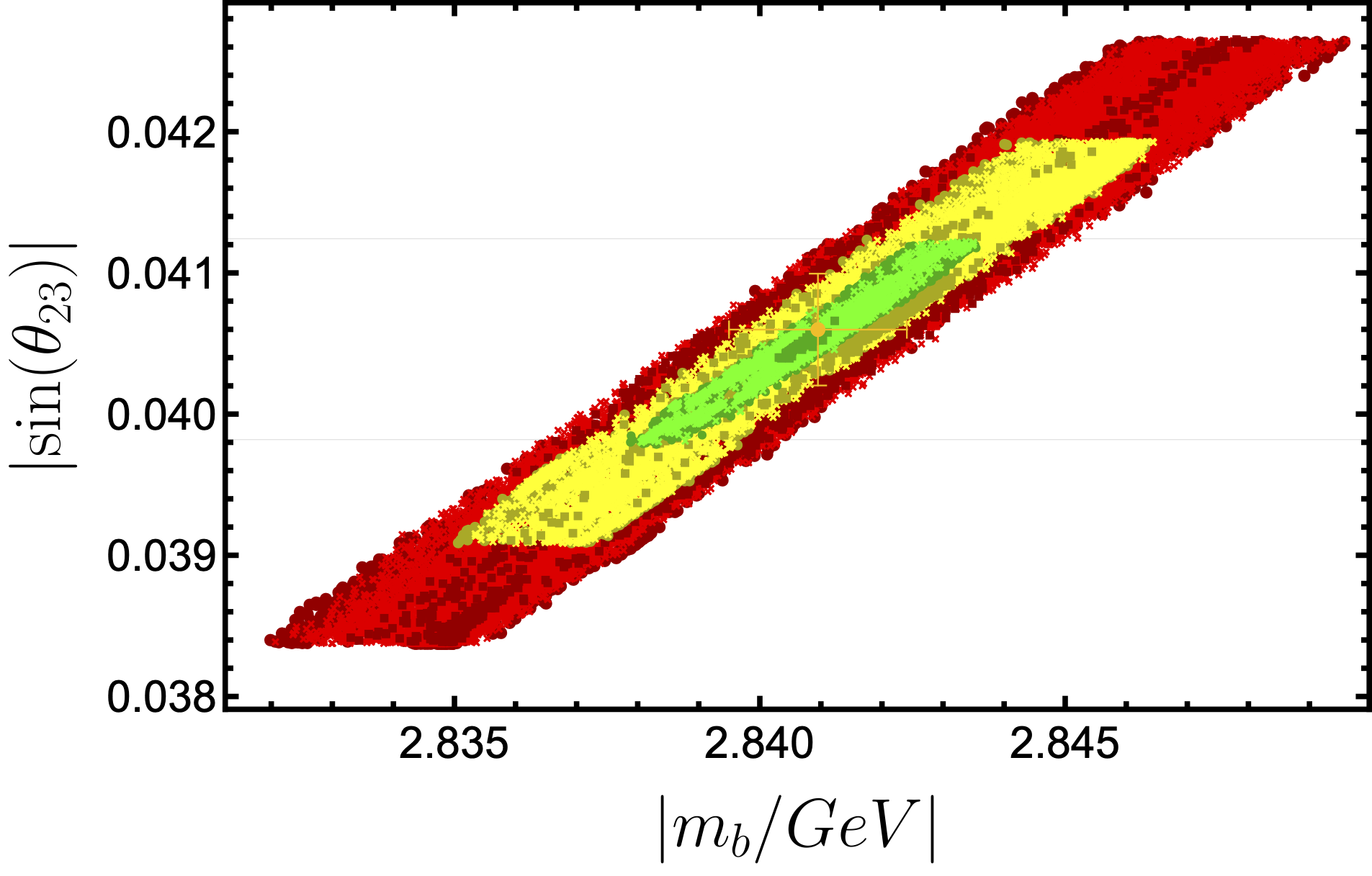}
        \caption{$m_b$ vs $\sin \left(\theta_{23}^{\text{\tiny{CKM}}}\right)$ global fit distribution graph.}
        \label{fig:mb_s23_B}
    \end{subfigure}
    \begin{subfigure}[t]{0.46\textwidth}
        \includegraphics[width=\textwidth]{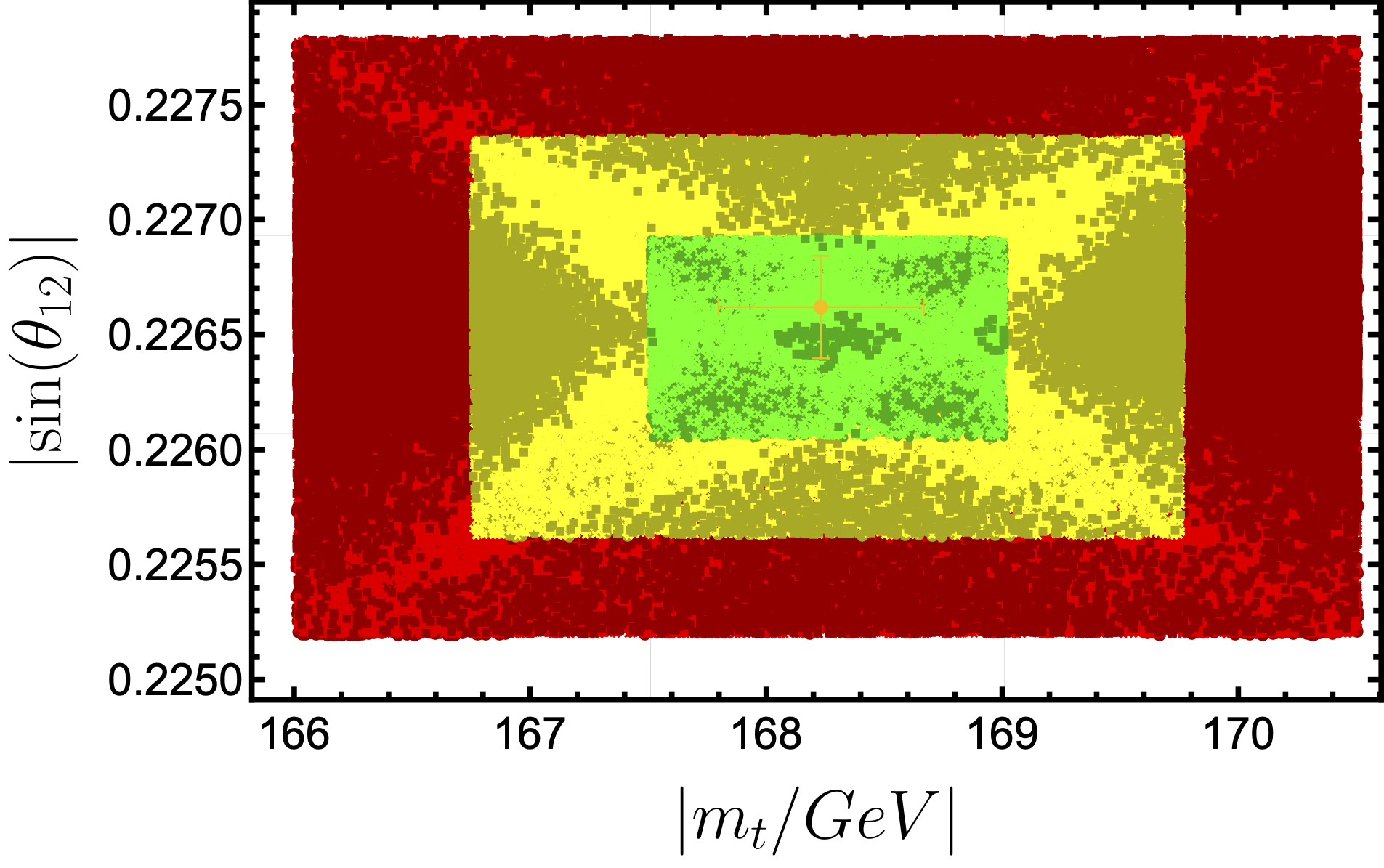}
        \caption{$m_t$ vs $\sin \left(\theta_{12}^{\text{\tiny{CKM}}}\right)$ global fit distribution graph.}
        \label{fig:mt_s12_B}
    \end{subfigure}
    \caption{Selected observable correlation plots. Grid lines represent areas of the experimental data with one standard deviation. Red, yellow, and green colors are used for the values $\sigma_{\text{max}}<3$, $2$, and $1$, respectively. Whereas, discs, crosses, and squares correspond to $\left\langle\sigma\right\rangle / \sigma_{\text{max}}$: $0.5-1.0$, $0.33-0.5$, $0-0.33$, respectively.}
    \label{fig:observ_corr_B}
\end{figure}

Plots in the Figs.~\ref{fig:md_ms_B},~\ref{fig:md_mb_B} and ~\ref{fig:ms_mb_B} demonstrate a direct-linear correlation between all three down sector quark masses. This is an immediate consequence of the fact that all three strongly depend on the $\beta_d$ input parameter, Figs.~\ref{fig:bd_md_B},~\ref{fig:bd_ms_B}, and ~\ref{fig:bd_mb_B}. A second local minimum of the best fit of the model in Fig.~\ref{fig:md_mu_B} is the same minimum that appeared in Figs.~\ref{fig:gd_md_B} and~\ref{fig:gu_mu_B}. 

Figs.~\ref{fig:md_s23_B},~\ref{fig:ms_s23_B} and ~\ref{fig:mb_s23_B} show the correlations between down quark sector masses and CKM mixing angle, $\sin \left(\theta_{23}^{\text{\tiny{CKM}}}\right)$. This can be seen immediately from the direct linear dependence of $\sin \left(\theta_{23}^{\text{\tiny{CKM}}}\right)$ on $\beta_d$ and $\gamma_d$, Fig.~\ref{fig:bd_s23_B} and Fig.~\ref{fig:gd_s23_B}, respectively. Furthermore, $m_d$, $m_s$, and $m_b$ all depend linearly on $\beta_d$ with various level of strength, Figs.~\ref{fig:bd_md_B},~\ref{fig:bd_ms_B}, and~\ref{fig:bd_mb_B}. Finally, in Fig.~\ref{fig:mt_s12_B} one can observe a "star"like pattern similar to the one given in the $\varepsilon$ vs $\sin \left(\theta_{12}^{\text{\tiny{CKM}}}\right)$ plot of Fig.~\ref{fig:eps_s123_B}. This similarity arises from the linear behaviour of $m_t$ on $\varepsilon$ in the limit $\varepsilon\rightarrow1$, eq.\eqref{eq:eps_limit_2}.

Following the phases assessment in the text succeeding the Eq.~\eqref{eq:MassM_uU}, in the situations when output observable variables were generated with a negative sign, the later was omitted.

\FloatBarrier
%
\section{Results}
\label{sec:results}
The results of the model predictions are given and elaborated on in the present section. This interpretation of 331 model, inspired by $SU(6)\otimes U(1)$, anticipates SM up and down quark masses, along with, CKM mixing angles for total of seven input parameters. Down, up and up type BSM isosinglet quark sectors are regulated by two parameters each. In addition,  light and massive BSM up quarks are mixed with an additional parameter denoted as $\varepsilon$.   The Tab.~\ref{tab:benchmark_param} provides collective list of the input parameters for the three most appropriate and significant benchmark points. The first benchmark point (BP1) is described as a point with the smallest $\chi^2$ of roughly $0.668$, which has maximum deviation from experimental results of $0.611\sigma$ ~Eq.~\eqref{eq:sigma_def}. 
The second benchmark point (BP2), contrasted with the first, is defined as the position in a parameter space scan with the lowest overall collection of deviations for all nine observable variables at present with a maximum deviation of $\sim0.487\sigma$.  Finally, we give the average of all data points obtained by $\forall \sigma_{\text{max}}\leq1$ as the third benchmark point (BP3), designated as BP3$_{\langle \rangle}$ in Tab .~\ref{tab:benchmark_param}, while the spread (error) of all points contributing to  $\forall \sigma_{\text{max}}\leq1$  is expressed as \emph{Spread}. The deviation $\sigma$ is described as follows

\begin{align}
    \label{eq:sigma_def}
    \sigma &= \left|\frac{x_{\text{exp}}-x_{\text{th}}}{x_{\text{err}}}\right|,
\end{align}
here $x$ indicates any of the observable variables from Tab.~\ref{tab:benchmark_obs}, \emph{exp.} stands for the experimentally obtained value, \emph{th} corresponds to the simulated observable value from the run of the parameter space scan, and lastly, \emph{err.} means the error for the experimentally obtained value.

\begin{table}[H]
    \centering
    {\footnotesize
    \begin{tabular}{cllll}
        \hline
        \text{par.} & $\text{BP1}$ & $\text{BP2}$ & $\text{BP3}_{\langle \rangle}$ & $\text{BP3}_{\text{spread}}$ \\ \hline
        $\beta_d$ & 0.0453747 & 0.04551359398035038 & 0.0455525 & 0.000216767 \\
        $\gamma_d$ & 0.00206769 & 0.0020552844447113165 & 0.00206312 & 0.0000139374 \\
        $\beta_u$ & -0.00741599 & -0.007358120617229136 & -0.00737406 & 0.0000430917 \\
        $\gamma_u$ & 0.000109067 & 0.00011100359211096111  & 0.00011017 & 0.00000139 \\
        $\beta_U$ & 0.0491484 & 0.04924143245126722 & 0.0492127 & 0.000405244 \\
        $\gamma_U$ & -0.0300187 & -0.030030668585101235 & -0.0300332 & 0.000259112 \\
        $\varepsilon$ & 1.01284 & 1.012851024748183 & 1.01268 & 0.00204506 \\
        \hline
    \end{tabular}
    }
    \caption{Model input parameters for the several benchmark points given in Tab.~\ref{tab:benchmark_obs}}
    \label{tab:benchmark_param}
\end{table}

\begin{figure}[H]
    \centering
    \includegraphics[width=0.8\textwidth]{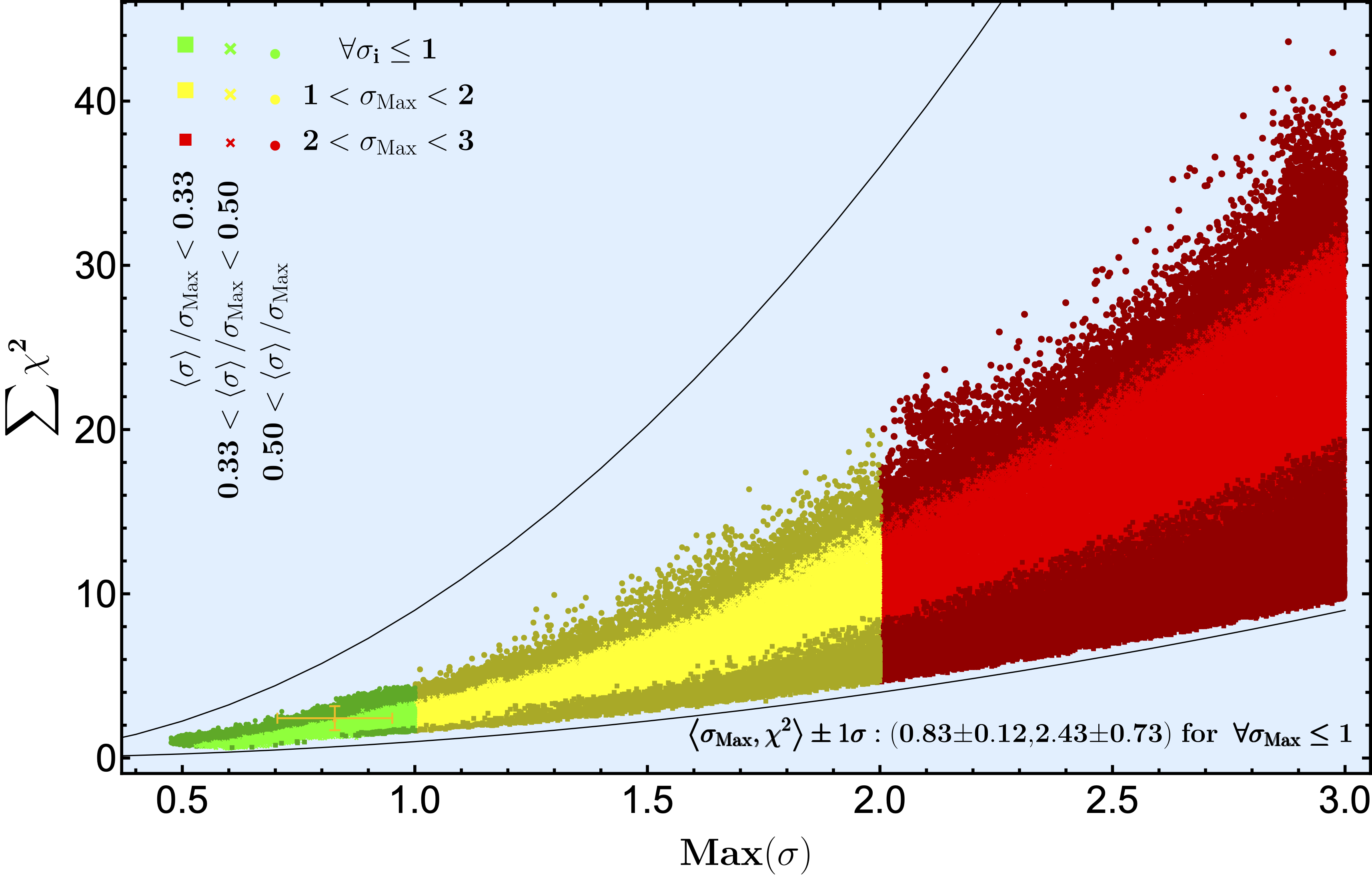}
    \caption{A plot of the distribution of model global fit vs maximum deviation (up to $3\sigma$). The theoretical upper and lower bounds are represented by solid curves, and the mean value is denoted by $\langle\rangle$.}
    \label{fig:max_sigma_chi2}
\end{figure}

Parameter scanning is very sensitive to the precision of input parameter values, so their values are given in Tab.~\ref{tab:benchmark_param} with up to twenty decimal places. The best result for $\chi^2$ for the sum of seven input parameters is given in columns 4 and 5 of the table.~\ref{tab:benchmark_obs} with a $\chi^2\approx 0.668$. As is observed, $m_{u}$ contributes the most to the $\chi^2$, but the third generation quark masses of up and down sectors generate a significantly lower imprecision to the $\chi^2$.  Then, as a result of finding the smallest combination of the deviations from the experimental values (2nd and 3rd columns of Tab.~\ref{tab:benchmark_obs}), the best obtained point is given in the 6th and 7th columns from Tab.~\ref{tab:benchmark_obs} with $\chi^2\approx 1.257$ and $\sigma_{\text{max}}\approx 0.487$. Finally, we collect all points with maximum deviations ($\sigma_{\text{max}}\leq1$) to generate mean and spread values for the set of the observable variables, given in the 8'th and 9'th column of Tab.~\ref{tab:benchmark_obs} with $\chi^2\approx 1.819$. These numbers represent the location and size of the region, with deviations from the experimental values smaller than one (green area in Fig.~\ref{fig:max_sigma_chi2}).

\begin{table}[H]
    \centering
    {\footnotesize
       \begin{tabular}{ccccccccc}
        \hline
        \scriptsize{\text{Observable}} & \multicolumn{2}{c}{\text{Experimental~\cite{Fritzsch:2021ipb}}} & \multicolumn{2}{c}{\text{BP1}} & \multicolumn{2}{c}{\text{BP2}} & \multicolumn{2}{c}{\text{BP3}} \\ \hline
        & \text{Value} & \text{Err.} & \text{Value} & $\sigma$ & \text{Value} & $\sigma$ & $\langle \rangle$ & Spread \\ \hline
        $m_d$ \scriptsize{(\text{MeV})} & 2.67 & 0.19 & 2.61 & 0.30 & 2.58 & 0.45 & 2.60 & 0.03 \\
        $m_s$ \scriptsize{(\text{MeV})} & 53.16 & 4.61 & 54.90 & 0.38 & 55.02 & 0.403 & 55.08 & 0.24 \\
        $m_b$ \scriptsize{(\text{GeV})} & 2.839 & 0.026 & 2.841 & 0.077 & 2.841 & 0.091 & 2.841 & 0.001 \\
        $m_u$ \scriptsize{(\text{MeV})} & 1.23 & 0.21 & 1.36 & 0.61 & 1.33 & 0.48 & 1.34 & 0.02 \\
        $m_c$ \scriptsize{(\text{MeV})} & 620 & 17 & 616 & 0.23 & 612 & 0.48 & 613 & 3.9 \\
        $m_t$ \scriptsize{(\text{GeV})} & 168.26 & 0.75 & 168.26 & 0.0017 & 168.42 & 0.21 & 168.27 & 0.44 \\
        $M_{U}$ \scriptsize{(\text{GeV})} & \multicolumn{2}{c}{---} & 3109 & - & 3109 & - & 3110 & 33 \\
        $M_{C}$ \scriptsize{(\text{GeV})} & \multicolumn{2}{c}{---} & 4296 & - & 4298 & - & 4298 & 31 \\
        $M_{T}$ \scriptsize{(\text{GeV})} & \multicolumn{2}{c}{---} & 83548 & - & 83555 & - & 83551 & 37 \\
        $\sin(\theta_{12})$ & 0.22650 & 0.000431 & 0.22651 & 0.020802 & 0.22666 & 0.36099 & 0.22654 & 0.000233 \\
        $\sin(\theta_{23})$ & 0.04053 & $^{+0.000821}_{-0.000601}$ & 0.04056 & 0.047517 & 0.04062 & 0.12329 & 0.04066 & 0.000156 \\
        $\sin(\theta_{13})$ & 0.00361 & $^{+0.000110}_{-0.000090}$ & 0.00360 & 0.063327 & 0.00366 & 0.48656 & 0.00364 & 0.000035 \\
        $\sin(\theta^{K^\pm}_{12})$ & \multicolumn{2}{c}{---} & 0.79602 & - & 0.79513 & - & 0.79544 & - \\
        $\sin(\theta^{K^\pm}_{23})$ & \multicolumn{2}{c}{---} & 0.01713 & - & 0.01710 & - & 0.01707 & - \\
        $\sin(\theta^{K^\pm}_{13})$ & \multicolumn{2}{c}{---} & 0.01232 & - & 0.01235 & - & 0.01238 & - \\
        $\sin(\theta^{K^{0}}_{12})$ & \multicolumn{2}{c}{---} & 0.63813 & - & 0.63688 & - & 0.63737 & - \\
        $\sin(\theta^{K^{0}}_{23})$ & \multicolumn{2}{c}{---} & 0.04605 & - & 0.04607 & - & 0.04607 & - \\
        $\sin(\theta^{K^{0}}_{13})$ & \multicolumn{2}{c}{---} & 0.01635 & - & 0.01635 & - & 0.01635 & - \\
        \hline
        $\chi^2$ & \multicolumn{2}{c}{---} & \multicolumn{2}{c}{$0.668$} & \multicolumn{2}{c}{$1.257$} & \multicolumn{2}{c}{$1.819$} \\
        \hline
        \end{tabular}
        }
   \caption{Various benchmark points of the model with the smallest $\chi^2$, the smallest $\sigma_{\text{max}}$, and mean value for $\forall \sigma_{\text{max}}\leq 1$; where $\sigma$ is the standard deviation and has no units, Eq.~\eqref{eq:sigma_def}. The obtained values shown above have been rounded to have the same significant numbers as the experimental results.}
    \label{tab:benchmark_obs}
\end{table}

The masses and mixing angles in Tab.~\ref{tab:benchmark_obs} were defined as eigenvalues of mass matrices in Eqs.~\eqref{eq:MassM_d},~\eqref{eq:MassM_uU}, and as in Eq.~\eqref{eq:CKMangles} for $V^W_{CKM}$(Eq.~\eqref{eq:CKM_SM}), $V^{K^{\pm}}$(Eq.~\eqref{eq:CKM_Kpm}), $V^{K^0}$(Eq.~\eqref{eq:CKM_K0}), respectively.

Figure~\ref{fig:max_sigma_chi2} summarizes all the data points collected during input parameter space scan according to two criteria: horizontal axis corresponds to $\sigma_{\text{max}}$ which represents the maximum deviation of each data point with respect to the experimental value obtained up to date, whereas the vertical axis shows the corresponding $\chi^2$ values for each data point obtained. The plot in Fig.~\ref{fig:max_sigma_chi2} is divided vertically into three horizontal regions according to the value of $\sigma_{\text{max}}$: $0-1$, $1-2$, $2-3$; vertical region is separated into three categories as well, according to the values of $\left\langle\sigma\right\rangle / \sigma_{\text{max}}$: $0-0.33$, $0.33-0.5$, $0.5-1.0$. The subdivision according to the last category represents the spread of all errors that contribute the total $\chi^2$. The solid curves on the plot stand for upper and lower theoretical limits for this plot given by $\chi^2=9\sigma_{\text{max}}^2$ and $\chi^2=\sigma_{\text{max}}^2$, respectively.

\FloatBarrier
%
\section{Discussion}
\label{sec:discussion}
The plots shown in the previous section, Figs.~\ref{fig:bg_observ_corr_B} and~\ref{fig:observ_corr_B}, can be used to identify and determine the reasons of varying levels of correlation between parameters and observable variables.  As expected, the $\gamma$ and $\beta$ parameters have an effect on the mass values of quarks in the up and down sectors. For example, $\gamma_u$ is expected to be strongly correlated with $m_u$, and $m_c$ and $m_t$ are expected to be weakly correlated. However, because $\beta u$ is about 70 times bigger than $\gamma_u$, the strong correlation of $\gamma_u$ with $m_u$ is blurred into medium level due to $\beta_u$ interference. $\beta_u$, as predicted, has a strong correlation with $m_c$ and $m_t$. It has a little effect on $m_u$ due to the relative size of $\beta_u$ in comparison to $\gamma_u$. The presence of BSM heavy isosinglet quarks (henceforth the \emph{BSM effect}), is another factor in determining the masses of the SM up sector quarks. This effect is governed in the model by the input parameter $\varepsilon$, which is the dominant influence parameter on the $m_t$ mass and in the limit $\varepsilon$ approaches one $m_t$ vanishes. 

On the other hand, the situation differs drastically for the down sector of the SM. $\gamma_d$ and $m_d$ are correlated on a medium level, much like the up sector. The leading effect of $\beta_d$ results in a subdominant correlation between $\gamma_d$ and the heavier down quark mass eigenvalues ($m_s$ and $m_b$). Correlation of $\beta_d$ with $m_d$, $m_s$, and $m_b$ is enhanced proportionally to the mass of the down sector quark, because the effect of $\gamma_d$ becomes less apparent with larger mass of the quark. Since there is no BSM effect in strong contrast with the up SM sector, lighter mass eigenvalues', \emph{e.g.} $m_d$, dependence is lead by $\gamma_d$, whereas larger mass eigenvalues', \emph{e.g.} $m_b$, dependence is dominated by $\beta_d$.

Regarding the CKM mixing angles, only the $\sin\left(\theta_{23}^{\text{\tiny{CKM}}}\right)$ has a correlation pattern with $\gamma_d$, $\beta_d$ input parameters and down sector SM quark masses. The correlation of other CKM mixing angles with input parameters or SM quark masses is much weaker or not observed at all.

As previously stated, CP violating phases are not taken into account in the present paper and left for consideration elsewhere. As a result, the elements of the mass matrices are selected to be real numbers.  Therefore, some of the eigenvalues of mass matrices and some elements of the CKM matrix are obtained as negative.  By adding phase multipliers to the democratic mass matrix elements, these negative signs can be removed and correct CP violating phases can be obtained. These multipliers are expected to help determine the values of  $\chi^2$ and  $\sigma_{\text{max}}$ as close to zero as possible. The effect of the phases on the quark masses and CKM mixing angles will be further investigated in the future.  
%
\section{Conclusion}
\label{sec:conclusion}
The utilization of the DMM technique to the quark sector of the $SU(6)$ symmetry motivated 331 model Variant-B is the subject of current work. Model stands out as one of the simplest extensions of SM. Using a total of ten parameters, the quark masses and mixing angles can be obtained within one standard deviation of the experimental values. A set of three parameters ($a, \gamma, \beta$) primarily control each quark sector (up, down and isosinglet up). In addition, one of the three parameters controls how SM and BSM isosinglet up type quarks are mixed. Therefore, all masses and mixing angles of SM and BSM isosinglet quarks are successfully predicted. There are a total of eighteen observable variables, nine of which are SM variables.

The best fit benchmark points are obtained by performing detailed analysis. It is found that, the best fit point is the point with the lowest $\chi^2=0.668$ and the maximum standard deviation $0.611$ from the experimental value, which is corresponding  to $m_u$. The other critical benchmark point is the one which has the lowest achievable error of standard deviation from the experimental data, with a $\chi^2=1.257$ value and the lowest maximum deviation of $0.487$. Besides producing the point with mean value for all created data set with the $\sigma_{\text{max}}\leq1$ condition, the plot summarizing all the data set in $\sigma_{\text{max}}$ vs $\chi^2$ graph is also generated.

In our previous paper, the democratic parameterization Variant-A of 331 model has succeeded in guiding us to the SM quark masses and hierarchy between them in accordance with the recent experimental data. Among the goals of this study, is to confirm that the parameterization at hand is valid for the Variant-B as well. The current work proves that SM quark masses and hierarchy among them are capable of being produced successfully via the democratic parameterization. Additionally, CKM mixing angles are also obtained within appropriate experimental limits. This leads us to a conclusion that further studies on altered parameter schemes based on fundamental democratic pattern are well motivated. Future research should examine UV models of flavor symmetry that naturally lead to democratic based scheme of quark mass sector. Conclusion drawn from the foregoing is that this may also provide a solution to the hierarchy problem.

%
\acknowledgments
RC was supported by Ege University Scientific Research Projects Coordination under Grant Number FGA-2021-22954. OP was supported by the Samsung Science and Technology Foundation under Grant No. SSTF-BA1602-04 and National Research Foundation of Korea under Grant Number 2018R1A2B6007000.
\appendix
%
%
\bibliographystyle{JHEP}
\bibliography{references}

\end{document}